\documentclass[aps,prx,floatfix,twocolumn,superscriptaddress,longbibliography]{revtex4-2}

\usepackage{graphicx}
\usepackage[english]{babel}
\usepackage{standalone}
\usepackage{amsthm}
\usepackage{thmtools, thm-restate}
\usepackage{amsfonts, amsmath, amssymb}
\usepackage{dcolumn}
\usepackage{bm}
\usepackage{dsfont}
\usepackage[utf8]{inputenc}
\usepackage[colorlinks,citecolor=blue,linkcolor=blue,urlcolor=blue]{hyperref}
\usepackage{physics}
\usepackage{mathtools}
\usepackage{float}
\usepackage{algorithm}
\usepackage{stmaryrd} 
\usepackage{enumerate}
\usepackage{quantikz}

\DeclarePairedDelimiterX{\mynorm}[1]{\lVert}{\rVert}{#1}

\newcommand{\expect}[1]{\mathbb{E}\left[#1\right]}
\newcommand{\expectt}[2]{\mathbb{E}_{#1}\left[#2\right]}

\def\d{{\rm{d}}}

\renewcommand{\tr}[1]{\mathrm{Tr}\left[#1\right]}

\newcommand{\rhohat}{\hat{\rho}}
\newcommand{\Uhat}{\hat{U}}

\newcommand{\ad}{a^{\dagger}}

\newcommand{\g}{\gamma}

\newcommand{\SO}[1]{\mathrm{SO}(#1)}
\newcommand{\Orth}[1]{\mathrm{O}(#1)}

\newcommand{\bsmu}{\boldsymbol{\mu}}
\newcommand{\bsnu}{\boldsymbol{\nu}}

\newcommand{\Pau}{\mathrm{P}_{n}}
\newcommand{\Cliff}{\mathrm{Cl}_{n}}
\newcommand{\Sym}{\mathrm{Sym}}
\newcommand{\Mn}{\mathrm{M}_{n}}

\newcommand{\Ccal}{\mathcal{C}}
\newcommand{\Hcal}{\mathcal{H}}
\newcommand{\Ecal}{\mathcal{E}}
\newcommand{\Rcal}{\mathcal{R}}

\newcommand{\Ucal}{\mathcal{U}}
\newcommand{\Ical}{\mathcal{I}}
\newcommand{\Gcal}{\mathcal{G}}
\newcommand{\Pcal}{\mathcal{P}}
\newcommand{\Mcal}{\mathcal{M}}
\newcommand{\Lcal}{\mathcal{L}}
\newcommand{\Scal}{\mathcal{S}}
\newcommand{\Zcal}{\mathcal{Z}}

\newcommand{\bea}{\begin{equation}\begin{aligned}}
\newcommand{\eea}{\end{aligned}\end{equation}}

\theoremstyle{definition}
\newtheorem{definition}{Definition}

\theoremstyle{plain}
\declaretheorem[name=Theorem]{thm}
\declaretheorem[name=Lemma]{lem}
\declaretheorem[name=Proposition]{prop}

\begin{document}

\title{Unified Framework for Matchgate Classical Shadows}

\author{Valentin Heyraud} 
\email{v.heyraud@instadeep.com}
\affiliation{InstaDeep, Paris, France}
\author{Héloise Chomet}
\affiliation{InstaDeep, London, United Kingdom}
\author{Jules Tilly}
\affiliation{InstaDeep, London, United Kingdom}

\begin{abstract}
Estimating quantum fermionic properties is a computationally difficult yet crucial task for the study of electronic systems. Recent developments have begun to address this challenge by introducing classical shadows protocols relying on sampling of Fermionic Gaussian Unitaries (FGUs): a class of transformations in fermionic space which can be conveniently mapped to matchgates circuits. The different protocols proposed in the literature use different sub-ensembles of the orthogonal group $\Orth{2n}$ to which FGUs can be associated. We propose an approach that unifies these different protocols, proving their equivalence, and deriving from it an optimal sampling scheme. We begin by demonstrating that the first three moments of the FGU ensemble associated with $\SO{2n}$ and of its intersection with the Clifford group are equal, generalizing a result known for $\Orth{2n}$ and addressing a question raised in previous works. Building on this proof, we establish the equivalence between the shadows protocols resulting from FGU ensembles analyzed in the literature. Finally, from our results, we propose a sampling scheme for a small sub-ensemble of matchgates circuits that is optimal in terms of number of gates and that inherits the performances guarantees of the previous ensembles.

\end{abstract}

\date{\today}
\maketitle

\section{Introduction}

Understanding the physics of correlated electronic systems is crucial for quantum chemistry~\cite{deglmannApplicationQuantumCalculations2015, williams-noonanFreeEnergyMethods2018, heifetzQuantumMechanicsDrug2020} and condensed matter physics~\cite{continentinoKeyMethodsConcepts2021, vandervenRechargeableAlkaliIonBattery2020}, with potential far-reaching applications in drug discovery~\cite{bluntPerspectiveCurrentStateoftheArt2022}, chemical engineering~\cite{lordiAdvancesOpportunitiesMaterials2021} and material science~\cite{caoPotentialQuantumComputing2018}. Simulating such many-body quantum systems on a classical computer is challenging due to the exponential scaling of the system's wave-function. This difficult task was one of the first applications envisioned for quantum computers~\cite{feynman1982, manin1980, ortizQuantumAlgorithmsFermionic2001, sommaSimulatingPhysicalPhenomena2002, sommaQuantumSimulationsPhysics2003} and it remains one of the most promising~\cite{verstraeteQuantumCircuitsStrongly2009,weckerSolvingStronglyCorrelated2015,jiangQuantumAlgorithmsSimulate2018,
smithSimulatingQuantumManybody2019,tacchinoQuantumComputersUniversal2020,
bharti2022,fausewehQuantumManybodySimulations2024}. However, current quantum devices are noisy and of limited size, which puts severe restrictions on the quantum algorithms that can be reliably executed. In that context, variational quantum algorithms~\cite{cerezoVariationalQuantumAlgorithms2021} have emerged as a popular class of algorithms addressing these hardware constraints. These versatile algorithms rely on a classical optimization of a variational ansatz obtained by applying a parameterized quantum circuit to some fixed initial state. Among these algorithms, the variational quantum eigensolver~\cite{peruzzoVariationalEigenvalueSolver2014, mccleanTheoryVariationalHybrid2016, tillyVariationalQuantumEigensolver2022} has attracted lot of attention in view of the simulation of many-body systems on near term devices. 

A critical step of this algorithm is the estimation of the expectation values of a set of observables. This operation occurs repeatedly during the optimization of the variational ansatz, which requires measuring the system Hamiltonian, and at the end of the algorithm to characterize the obtained quantum state. To this end, typical quantities of interest are the $k$-body reduced density matrices ($k$-RDM), which encode many relevant physical properties of the simulated systems~\cite{colemanReducedHamiltonianOrbitals1980}. For the study of fermionic systems on quantum devices, the 2-RDM can be used to estimate the system energy~\cite{rubinApplicationFermionicMarginal2018, tillyReducedDensityMatrix2021}, the associated gradients~\cite{obrienCalculatingEnergyDerivatives2019, overyUnbiasedReducedDensity2014} and multipole moments~\cite{gidofalviMolecularPropertiesVariational2007}; while the 3-RDM finds useful applications for condensed matter models~\cite{tsuneyukiTranscorrelatedMethodAnother2008, petersonMoreRealisticHamiltonians2013}. Multiple methods have been developed to efficiently estimate fermionic $k$-RDM. \citet{bonet-monroigNearlyOptimalMeasurement2020} proposed a strategy based on the gathering of the target observables into cliques of commuting operators that can be measured simultaneously. Although nearly optimal for the 1- and 2-RDMs, the proposed methods cannot be easily generalized to larger values of $k$. Another similar strategy has been proposed in Ref.~\cite{jiangOptimalFermiontoqubitMapping2020} but requires the use of multiple ancillary qubits.

Recognizing these limitations, \citet{zhaoFermionicPartialTomography2021} proposed a probabilistic strategy, building on the recent breakthrough of the classical shadows protocol \cite{huangPredictingManyProperties2020, elbenRandomizedMeasurementToolbox2023}. This protocol consists in the construction of a classical representation of a quantum state obtained from the measurement performed on the system evolved according to a random unitary transformation. This representation is then used to build estimators of the quantities of interest. A key choice in this procedure is the ensemble of random unitary transformations used to derive the classical representation. The strategy proposed in Ref.~\cite{zhaoFermionicPartialTomography2021} relies on a subgroup of the group of Fermionic Gaussian Unitaries (FGUs), which is a particular subset of the transformations preserving the linear span of the $2n$ Majorana operators associated with $n$ fermionic modes (see Sec.~\ref{subsec:FGU_and_matchgates} for more details). Under the Jordan-Wigner mapping~\cite{jordanUeberPaulischeAequivalenzverbot1928}, these transformations correspond to matchgate circuits~\cite{terhalClassicalSimulationNoninteractingfermion2002}, a class of efficiently classically simulable circuits generated by specific Pauli rotations acting on adjacent pairs of a qubits chain~\cite{valiantQuantumComputersThat2001a,jozsaMatchgatesClassicalSimulation2008}. Up to a global phase, FGUs are in one-to-one correspondence with the elements of the group of special orthogonal matrices $\mathrm{SO}(2n)$. This allows for efficient classical simulation schemes that do not rely explicitly on the previous mapping~\cite{knillFermionicLinearOptics2001, divincenzoFermionicLinearOptics2005, bravyiLagrangianRepresentationFermionic2005, caiTheoryMatchgateComputations2009}, which in turns enables an efficient classical post-processing for the classical shadows protocol. 

\citet{zhaoFermionicPartialTomography2021} considered the subset of FGUs belonging to the Clifford group (under an arbitrary fermion-to-qubit mapping). The obtained ensemble corresponds to a subgroup of $\mathrm{SO}(2n)$ solely composed of signed permutation matrices. Using this ensemble, they derived an asymptotically optimal classical shadows protocol for the estimation of fermionic $k$-RDMs. A limitation of the proposed method lies in the use of the subgroup of Clifford FGUs, which appears to single out a preferred basis of Majorana operators. Following this work, \citet{wanMatchgateShadowsFermionic2023} introduced a classical shadows protocol using a larger ensemble, which we refer to as the generalized FGUs, in view of an application to hybrid quantum-classical quantum Monte-Carlo simulations (QC-QMC) of fermionic systems~\cite{hugginsUnbiasingFermionicQuantum2022}. Similarly to FGUs, generalized FGUs are in one-to-one correspondence with the orthonormal group $\mathrm{O}(2n)$ and correspond to matchgate circuits complemented with single-qubit Pauli gates under the Jordan-Wigner map. The authors prove that the first three moments of the uniform distributions on the generalized FGUs and on the subset of Clifford generalized FGUs are equal. This result enables them to simultaneously exploit the symmetries of the Clifford group and the invariance under rotation of the Haar measure on $\Orth{2n}$, thereby avoiding singling out a preferred basis of Majorana operators. Leveraging this finding, they show that their scheme is efficient in estimating various quantities such as the overlap between an arbitrary pure state and a fermionic Gaussian state. In a related work, \citet{ogormanFermionicTomographyLearning2022} provide a simplified analysis of the scheme presented in Ref.~\cite{zhaoFermionicPartialTomography2021}, offering a corrected expression of the estimators variances within and showing the efficiency of the Clifford FGU ensemble for the classical shadows estimation of various quantities. They also considered a shadows protocol associated with a specific subgroup of permutation corresponding to the so-called perfect-matchings, and proved a partial equivalence with the results of Ref.~\cite{zhaoFermionicPartialTomography2021}.

As noted by \citet{zhaoGrouptheoreticErrorMitigation2023}, so far the group of FGU corresponding to the continuous matrix group $\SO{2n}$ has not yet been analyzed in the context of classical shadows, and \citet{wanMatchgateShadowsFermionic2023} left the possibility to extend the matchgate 3-design property to this ensemble as an open question. Furthermore, to this day, the link between the shadows protocols corresponding to different sub-ensembles of FGU remains unclear. In this paper, we investigate the use of the FGU ensemble associated with $\SO{2n}$ for the classical shadows protocol. First, we consider the larger class of FGU ensembles that can be decomposed into ensembles of matchgate circuits with independent random Pauli rotations and show that under mild symmetry assumptions on the distributions of the random angles, these ensembles admit the same first three moments as their sub-ensembles belonging to the Clifford group. We refer to ensembles with this property as Clifford-3-cubatures. We find that the FGU ensemble associated with $\mathrm{SO}(2n)$ is a Clifford-3-cubature, thereby providing a positive answer to the open question introduced in Ref.~\cite{wanMatchgateShadowsFermionic2023}. Our result proves that the $\SO{2n}$ group leads to a shadows protocol that is equivalent to the one considered in Ref.~\cite{zhaoFermionicPartialTomography2021}. Second, we show that classical shadows protocols using ensembles of Clifford FGUs are unaffected by the injection of reflections with respect to Majorana operators. Precisely, we complete the results of Ref.~\cite{zhaoFermionicPartialTomography2021} and show that the variances of the shadow estimators are invariant under such reflections. We also extend the results of Ref.~\cite{ogormanFermionicTomographyLearning2022} and show that ensembles of FGU which permutations correspond to the same perfect-matching also lead to shadow estimators with the same variances. This allow us to rigorously prove that the different FGU ensembles considered in the literature (in particular in Refs.~\cite{zhaoFermionicPartialTomography2021},\cite{wanMatchgateShadowsFermionic2023} and \cite{ogormanFermionicTomographyLearning2022}) yield equivalent classical shadows protocols. Finally, we present and discuss new sampling schemes for different subsets of Clifford FGU. Building on the previous equivalence result, we derive a sampling scheme that is optimal in terms of number of gates and that generates a FGU ensemble inheriting the best performance guarantees of the previous ensembles.

\section{Background}
\label{sec:background}

In this section we provide an overview of the background material necessary to develop and prove our results. First, we introduce some general notations and mathematical facts in Sec.~\ref{subsec:general_notations}. Then, we review the classical shadows protocol in Sec.~\ref{subsec:classical_shadows}. We gather useful definitions and results related to fermionic Gaussian unitaries and matchgate circuits in Sec.~\ref{subsec:FGU_and_matchgates}. Finally, we review existing results related to shadows protocols based on FGU ensembles in Sec.~\ref{subsec:previous-results}.

\subsection{Notations and mathematical preliminaries}
\label{subsec:general_notations}

Throughout this paper, we will denote $\mathbb{N}$ the set of non-negative integers, $\left[k\right] := \left\{1,\dots,k\right\}$ and we write ${\mathbb{N}^{\ast}:=\mathbb{N}\backslash\{0\}}$. For 
multi-indices $\bsmu,\bsnu\subset \mathbb{N}^{\ast}$ we write $\delta_{\bsmu\bsnu}$ the generalized Kronecker symbol that is equal to $1$ if $\bsmu=\bsnu$ and $0$ otherwise. We follow Ref.~\cite{zhaoFermionicPartialTomography2021} and write $\Sym(2n)$ the symmetric group on $2n$ objects. This group is faithfully represented by the group of $2n\times 2n$ permutation matrices, namely matrices which columns and rows have only one non-zero entry equal to 1. We denote $\Sym(2,2n)$~\footnote{The group $\Sym(2,2n)$ of generalized permutation matrices is denoted $\mathrm{B}(2n)$ in Ref.~\cite{wanMatchgateShadowsFermionic2023}} the generalized symmetric group of cyclic order $2$, which is defined as the wreath product ${\mathbb{Z}_{2} \wr \Sym(2n)}$. The group $\Sym(2,2n)$ is faithfully represented by permutation matrices whose non-zero entries take values in $\{-1, 1\}$. Every such matrix $M$ can be written $M=DP$ with $P$ a permutation matrix and $D$ a diagonal matrix with entries in $\{-1, 1\}$. The matrices $D$ can be seen as a faithful representation of $\mathbb{Z}^{n}_{2} \cong \{-1,1\}^{n}$ and for two generalized permutations matrices $M_{1}, M_{2}$ we have the semi-direct product $M_{1} M_{2} = (D_{1} P_{1} D_{2} P^{-1}_{1})(P_{1} P_{2} )$, so that ${\Sym(2,2n) = \mathbb{Z}^{n}_{2} \rtimes \Sym(2n)}$. We also write $\Sym^{+}(2n)$~\footnote{As the determinant of a permutation matrix is equal to the signature of the corresponding permutation, $\Sym^{+}(2n)$ is exactly the alternating group, i.e. the subgroup of even parity permutations denoted $\mathrm{Alt}(2n)$ in Ref.~\cite{zhaoFermionicPartialTomography2021}.} (respectively $\Sym^{+}(2, 2n)$ the subgroup of $\Sym(2n)$ (respectively $\Sym(2, 2n)$) corresponding to matrices with determinant $+1$. In the following we do not distinguish the previous groups and their matrix representations.

The Hilbert space of a system of $n$ qubits is denoted $\Hcal_{n}\cong \mathbb{C}^{2^n}$. Since they have the same dimension, $\Hcal_{n}$ is unitarily isomorphic to the state-space of a system $n$ fermionic modes, and upon fixing a fermion-to-qubits mapping we can identify both. Denote $\mathrm{U}(\Hcal_{n})$ the set of unitary operators on $\Hcal_{n}$. Operators acting trivially on all but the $k$-th qubit are written with a lower index $k$. In particular, we denote $X_k, Y_k, Z_k$ the Pauli operators associated with the $k$-th qubit. The identity operator is denoted $I$. $\Hcal_{n}$ is equipped with the usual canonical basis of eigenstates of the Pauli-$Z$ operators $\left\{\ket{z},\,z\in\left\{0,1\right\}^{n}\right\}$. We denote $\Pau$ the Pauli group, which contains all Pauli words of the form ${P=\lambda\prod_{k=1}^{n}P_k}$ with $P_k\in\left\{I, X_k, Y_k, Z_k\right\}$ and ${\lambda\in\left\{1,-1,i,-i\right\}}$. We write $\mathrm{Cl}_n$ for the Clifford group, which is defined as the group of unitary operators normalizing the Pauli group, namely ${\Cliff := \left\{C\in\mathrm{U}(\Hcal_{n})\,\vert\,C\Pau C^{\dagger}\subseteq \Pau\right\}}$. Recall that up to global phases $\Cliff$ is generated by the control-$Z$, the Hadamard and the phase gates, respectively denoted $CZ$, $H$ and $S$.

We write $\Lcal(V)$ the space of linear operators on a complex vector space $V$. The set $\Lcal(\Hcal_{n})$ is itself a Hilbert space equipped with the Hilbert-Schmidt inner product $\langle A,B\rangle_{HS} := \tr{A^{\dagger}B}$. We write ${\Lcal(\Lcal(\Hcal_{n}))}$ the vector space of superoperators on ${\Lcal(\Hcal_{n})}$ and we call a quantum channel any superoperator that is completely positive and trace preserving~\cite{watrousTheoryQuantumInformation2018}. In the following, superoperators will be denoted with calligraphic letters. In particular, for a unitary transformation $U$, we will write $\Ucal$ the corresponding unitary quantum channel defined as $\Ucal(A) := U^{\dagger} A U$ for $A\in \Lcal(\Hcal_{n})$.

\subsection{Classical shadows protocol}
\label{subsec:classical_shadows}

In this subsection we briefly review the classical shadows procedure introduced by Huang \textit{et al.} in Ref.~\cite{huangPredictingManyProperties2020}. The aim of this protocol is to estimate the expectation values of a set of $M$ observables $\left\{O_1,\dots, O_M\right\}\in\Lcal(\Hcal_{n})$ with respect to an unknown quantum state $\rho$ of which we have multiple copies. The first step of the procedure is to chose a unitary ensemble $\mathbb{U}$ characterized by a probability measure $\eta$ over $\mathrm{U}(\Hcal_{n})$ (or some subset thereof) which can be efficiently sampled. For each copy of $\rho$, one then draw a random unitary transformation $\Uhat\sim\mathbb{U}$, apply it to $\rho$ and perform a measurement of the resulting state in the computational basis. One then obtain a classical bit-string $\hat{z}$, whose probability distribution conditioned on $\Uhat$ is given by Born's rule
\bea
\mathbb{P}\left[\hat{z} = z\,\vert\, \Uhat\right] = \bra{z}\Uhat\rho\Uhat^{\dagger}\ket{z}\,.
\eea
Applying the inverse unitary $\Uhat^{\dagger}$ to the state $\ket{\hat{z}}$ and averaging over the realisations yields a mixed state that can be seen as the image of $\rho$ under the so-called measurement quantum channel
\bea
\mathcal{M}(\rho) &:= \expectt{\Uhat, \hat{z}}{\Uhat^{\dagger}\dyad{\hat{z}}\Uhat}\\
&\phantom{:}= \expectt{\Uhat}{\sum_{z\in\left\{0,1\right\}^{n}} \bra{z} \Uhat\rho \Uhat^{\dagger}\ket{z} \Uhat^{\dagger}\dyad{z} \Uhat}\,,
\eea
Assuming that $\mathcal{M}$ is invertible, one can define an estimator of $\rho$ as follow
\bea
\rhohat := \mathcal{M}^{-1}\left(\Uhat^{\dagger}\dyad{\hat{z}}\Uhat\right)\,.
\eea
This estimator and its realisations are referred to as classical shadows in the literature. Note that the requirement that $\Mcal$ must admit an inverse on the whole space $\Lcal(\Hcal_{n})$ can be relaxed to $\Mcal$ admitting an inverse on a subspace of $\Lcal(\Hcal_{n})$, provided that both $\rho$ and $U^{\dagger}\dyad{z}U$ belong to this subspace for any $z\in\left\{0,1\right\}^{n}$ and ${U\in\mathbb{U}}$~\cite{wanMatchgateShadowsFermionic2023}. 

Using the classical shadow $\rhohat$, one can then build estimators for the expectation values $o_i := \tr{O_i \rho}$ for $i\in [1,M]$ by defining
\bea
\hat{o}_i := \tr{O_i\rhohat}\,.
\eea
Remark that these estimators are to be computed classically from the knowledge of the sampled unitaries $\hat{U}$ and measurement outcomes $\hat{z}$. By construction, the classical shadows and the corresponding estimators are unbiased. In particular, the only part of the variance of $\hat{o}_i$ affected by the choice of $\mathbb{U}$ is the raw second moment $\expect{\hat{o}_i^{2}}$, which is also a simple majorant of $\mathrm{Var}\left[\hat{o}_i\right]$. \citet{huangPredictingManyProperties2020} also introduced another useful majorant of the previous variance, the so-called shadow norm which is defined as the supremum of the second raw moment over all possible states $\rho$. Using a median-of-mean estimator and the associated concentration inequalities~\cite{lerasleLectureNotesSelected2019, huangPredictingManyProperties2020}, it can be shown that a sample size
\bea
N \propto \frac{1}{\epsilon^{2}}\log\left(\frac{M}{\delta}\right)\max_{1\leq i\leq M}\bigl(\mathrm{Var}\left[\hat{o}_i\right]\bigr)
\label{eq:sample_size_scaling_cs}
\eea
is sufficient to estimate all the $M$ expectation values up to an error $\epsilon$ and with a probability of failure $\delta$. Ref.~\cite{huangPredictingManyProperties2020} motivates the use of a median-of-mean estimator by the obtention of a logarithmic scaling in both $M$ and $\delta$.
Depending on the observables to measure and on the unitary ensemble considered, the observables estimators might be bounded. In that case, a direct application of the Hoeffding inequality~\cite{hoeffdingProbabilityInequalitiesSums1994} shows that a simple mean-of-sample estimator can be used, leading to a similar scaling for $N$ (upon replacing the variances by the range of the estimators in Eq.~\eqref{eq:sample_size_scaling_cs}). However, as the inverse of the measurement channel might not be a quantum channel itself, the classical shadows are not well defined quantum states in general. In particular, $\rhohat$ might fail to be positive, which can make it difficult to bound for $\tr{O_i\rhohat}$.

In order to better appreciate the role of the chosen unitary ensemble in the classical shadows protocol, it is insightful to introduce the $t$-fold twirl~\cite{kohClassicalShadowsNoise2022} (or simply $t$-fold~\cite{roberts2017}) channel of $\mathbb{U}$, whose action on ${A\in\Lcal(\Hcal_{n}^{\otimes t})}$ is given by
\bea
\Ecal^{(t)}(A) &:= \int_{\mathrm{U}(\Hcal_{n})} (U^{\dagger})^{\otimes t} A U^{\otimes t} \eta(\d U)
\label{eq:t-fold-channel}
\eea
with $\eta$ the probability measure defining $\mathbb{U}$. The $t$-fold channel completely characterizes the first $t$ moments of the distribution $\eta$. Random unitary ensembles whose $t$-fold channel matches the $t$-fold channel of the Haar measure~\cite{meleIntroductionHaarMeasure2024} on $\mathrm{U}(\Hcal_{n})$ are said to be unitary $t$-designs. As an example, it is well known that the Clifford group $\Cliff$ equipped with the uniform distribution is a $3$-design although it fails to be a $4$-design~\cite{webbCliffordGroupForms2016a, zhuCliffordGroupFails2016}. The notion of $t$-design have also been extended to other unitary groups (see e.g. Ref.~\cite{mitsuhashiCliffordGroupUnitary2023}) and to continuous variable systems in Ref.~\cite{iosueContinuousVariableQuantumState2024a}. 
These ensembles have found numerous useful applications in quantum information theory~\cite{grossEvenlyDistributedUnitaries2007, harrow2009, royUnitaryDesignsCodes2009, roberts2017, nakajiExpressibilityAlternatingLayered2021, haferkamp2022, holmes2022}. In particular, the use of the uniform Clifford ensemble for the classical shadows protocol was investigated in Ref.~\cite{huangPredictingManyProperties2020}.

Having defined the $t$-fold channel of $\mathbb{U}$, one can use the linearity of both the trace and the expectation to rewrite the measurement channel as 
\bea
\Mcal(\rho) = \sum_{z\in\left\{0,1\right\}^n}\Tr_{1}\left[\Ecal^{(2)}(\dyad{z}^{\otimes 2})(\rho\otimes I)\right]\,,
\eea
where $\Tr_1$ denotes the partial trace over the first tensor component. Likewise, using the fact that $\tr{\Mcal^{-1}(A)B} = \tr{A\Mcal^{-1}(B)}$, one can write the second raw moment $\expect{\hat{o}_i^{2}}$ as follow
\bea
\sum_{z\in\{0,1\}^{n}}\tr{\Ecal^{(3)}(\dyad{z}^{\otimes 3})(\rho\otimes\Mcal^{-1}(O_i)\otimes\Mcal^{-1}(O_i))}.
\label{eq:second-raw-moment}
\eea
These expressions show that unitary ensembles sharing the same 3-fold channel yield classical shadows protocols whose efficiencies are essentially equal. As for the results presented in Ref.~\cite{wanMatchgateShadowsFermionic2023}, many of our results will rely on this observation.

\subsection{Fermionic Gaussian Unitaries and Matchgate Circuits}
\label{subsec:FGU_and_matchgates}

In the next paragraphs, we introduce some notations and definitions related to many-body fermionic systems. Then, we review the various ensembles of fermionic Gaussian unitaries used in the literature and give some of the associated results in the context of classical shadows.

Consider a system of $n$ fermionic modes whose creation and annihilation operators are denoted ${a_i,\,a^{\dagger}_i}$ with ${i\in\left[n\right]}$. The fermionic modes can equivalently be represented by the $2n$ Majorana operators
\bea
\g_{2p-1} := a^{\dagger}_p+a_p\,,\quad
\g_{2p} := i\left(a^{\dagger}_p-a_p\right)\,,
\eea
which are self-adjoint and satisfy the anti-commutation relations $\left\{\g_{k},\g_{l}\right\}=2\delta_{kl} I$. In the following we will denote $\binom{\left[2n\right]}{k}$ the set of subsets of $\left[2n\right]$ with $k$ elements, and identify an element $\bsmu\in\binom{\left[2n\right]}{k}$ with the corresponding $k$-multi-index ${\bsmu := (\mu_1, \dots, \mu_{k})}$ which elements are sorted in increasing order ${1\leq\mu_1<\dots<\mu_{k}\leq 2n}$. For any such $\bsmu$, define the associated $k$-degree Majorana operator
\bea
\g_{\bsmu} := \g_{\mu_1}\dots\g_{\mu_{k}}\,.
\eea
These operators form an orthogonal family for the Hilbert-Schmidt inner product.

The $k$-RDM of a state $\rho$ is defined as the tensor of order $2k$ whose entries are written
\bea
\prescript{k}{}D^{p_1\dots p_k}_{q_1\dots q_k}:=\tr{\ad_{p_1}\dots\ad_{p_k}a_{q_1}\dots a_{q_k}\rho}\,.
\eea
In order to estimate the $k$-RDM using a quantum computer, one needs to map the system of $n$ fermionic modes to a system of qubits. For the sake of clarity, we will use the Jordan-Wigner (JW) transformation throughout this paper, which we recall in Appendix~\ref{app:jordan-wigner}.
Having fixed a fermion-to-qubit mapping, we will no longer distinguish the fermionic from the qubit transformations. Note that the results presented in this paper are independent of the exact mapping used, as long as Majorana operators are mapped to Pauli ones.

\citet{zhaoFermionicPartialTomography2021} proposed a classical shadows protocol tailored for the estimation of fermionic $k$-RDMs relying on a subset of the continuous group of fermionic Gaussian unitary transformations. A FGU is a unitary transformation $U_{Q}\in\Lcal(\Hcal_{n})$ associated with some ${Q\in\SO{2n}}$ such that
\bea
U^{\dagger}_Q\g_{k}U_Q=\sum_{k=1}^{2n}Q_{kl}\g_{l}\,
\label{eq:adjoint_action_fgu}
\eea
for every $k\in\left[2n\right]$. As before, we denote $\Ucal_{Q}$ the corresponding unitary quantum channel. Using the Leibnitz formula for determinants and the commutation relation of the Majorana operators, we have
\bea
U^{\dagger}_Q\g_{\bsmu}U_Q=\sum_{\bsnu\in\binom{\left[2n\right]}{\abs{\bsmu}}}\det\left(Q_{\bsmu\bsnu}\right)\g_{\bsnu}\,,
\label{eq:fgu_on_majorana_products}
\eea
where $Q_{\bsmu\bsnu}\in\mathbb{R}^{\abs{\bsmu}\times\abs{\bsmu}}$ is defined by $(Q_{\bsmu\bsnu})_{ij} = Q_{\mu_i \nu_j}$ for $1\leq i,j \leq \abs{\bsmu}=\abs{\bsnu}$. Hence, every FGU is characterized up to a global phase by its associated element $Q\in\SO{2n}$. In particular, the map $Q\mapsto \Ucal_{Q}$ yields a faithful representation of $\SO{2n}$ on $\Lcal(\Hcal_{n})$ that satisfies
\bea
\Ucal_{Q'}\circ \Ucal_{Q} = \Ucal_{QQ'}
\label{eq:composition_fgu}
\eea
for all $Q,Q'\in\SO{2n}$. Using the Cauchy-Binet formula~\cite{taoTopicsRandomMatrix2012a} and the matrix elements ${\tr{\g_{\bsnu}^{\dagger}\Ucal_{Q}(\g_{\bsmu})}}$ in Liouville representation, one can show that this representation is orthogonal~\cite{zhaoFermionicPartialTomography2021}.

The existence of an homomorphism from $\SO{2n}$ to $\Lcal(\Lcal(\Hcal_{n}))$ allows to transform a decomposition of the element $Q$ in elementary building blocks into a decomposition of the corresponding $U_{Q}$. Of note are the decompositions in terms of Givens rotations~\cite{givensComputationPlainUnitary1958} which are rotations in planes spanned by two coordinate axes of $\mathbb{R}^{2n}$. More precisely, a Givens rotation $g_{ij}(\theta)$ of axes $i,j \in\llbracket 1,2n\rrbracket$ and angle $\theta\in(-\pi,\pi]$ is defined as the rotation whose matrix in the canonical basis of $\mathbb{R}^{2n}$ reads
\bea
\bordermatrix{         &        &        &    i   &        &    j   &       \\ 
            &      1   & \cdots &    0   & \cdots &    0   & \cdots &    0   \\
            &  \vdots & \ddots & \vdots &        & \vdots &        & \vdots \\
        i   &     0   & \cdots &    \cos(\theta)   & \cdots &   -\sin(\theta)   & \cdots &    0   \\
            &  \vdots &        & \vdots & \ddots & \vdots &        & \vdots \\
        j   &     0   & \cdots &    \sin(\theta)   & \cdots &    \cos(\theta)   & \cdots &    0   \\
            &  \vdots &        & \vdots &        & \vdots & \ddots & \vdots \\
            &     0   & \cdots &    0   & \cdots &    0   & \cdots &    1
}\,.
\eea
Defining
\bea
G_{ij}(\theta) = \exp\left(-\frac{\theta}{2}\g_{i}\g_{j}\right)\,,
\eea
and writting $\Gcal_{ij}(\theta)$ the corresponding channel, we have
\bea
\Gcal_{ij}(\theta)(\g_{k}) &:= G_{ij}(\theta)^{\dagger}\g_{k}G_{ij}(\theta)\\
&\phantom{:}= \sum_{l=1}^{2n} \left[g_{ij}(\theta)\right]_{kl}\g_{l}\,,
\eea
that is $g_{ij}(\theta)$ is represented by $\Gcal_{ij}(\theta)$. From what precedes, decomposing $Q\in\SO{2n}$ as a product of $g_{ij}(\theta)$ yields a decomposition of $\Ucal_{Q}$ as a composition of $\Gcal_{ij}(\theta)$. Such decompositions have found useful applications in the literature related to FGU and the simulation of fermionic systems~\cite{weckerSolvingStronglyCorrelated2015, jiangQuantumAlgorithmsSimulate2018, kivlichanQuantumSimulationElectronic2018a}. 

Under the JW mapping, the FGUs correspond to the so-called matchgate circuits. A matchgate is defined as a two-qubits Pauli rotation of the form 
\bea
U(\theta) &= \exp\left( i\frac{\theta}{2}P\otimes P'\right)\,,\\
P\otimes P' &\in \left\{Z\otimes I,\, I\otimes Z,\, X\otimes X\right\}\,.
\eea
Considering that the $n$ qubits are placed on a line, matchgate circuits are then defined as the quantum circuits composed of matchgates acting on pairs of adjacent qubits. This correspondence between FGU and matchgate circuits, which was first proven in Ref.~\cite{knillFermionicLinearOptics2001}, can be elucidated as follow. First, notice that
\bea
i Z_{k} = \g_{2k-1}\g_{2k},\quad i X_{k}X_{k+1}=\g_{2k}\g_{2k+1}\,,
\eea
so that matchgates correspond to Givens rotations on adjacent qubits. In the following we write $g_{k}(\theta):= g_{k-1k}(\theta)$ these Givens rotations, and likewise for $G_{k}(\theta)$ and $\Gcal_{k}(\theta)$ for $2\leq k \leq 2n$. Then, remark that any elements of $\SO{2n}$ can be decomposed as a product of $g_{k}(\theta)$. We review an important scheme introduced in Ref.~\cite{hurwitzUberErzeugungInvarianten1897} that achieves such a decomposition in the next section.

The classical shadows introduced in Ref.~\cite{zhaoFermionicPartialTomography2021} are built using the subgroup of FGU belonging to the Clifford group. Considering a fermion-to-qubit transformation mapping the Pauli operators to the Majorana ones, a FGU is in the Clifford group if and only if its matrix $Q$ belongs to $\Sym^{+}(2,2n)$. The corresponding matchgate circuits are generated by matchgates with angles belonging to $\left\{0,\pi, \frac{\pi}{2}, -\frac{\pi}{2}\right\}$, which we refer to as the Clifford angles. To see this, remark that any rotation generated by a Pauli string can be transformed into a single-qubit $Z$-rotation upon conjugating with the adequate Clifford gates, and that the only $Z$-rotations in the Clifford group are the ones for which $\theta$ belongs to the Clifford angles (the angles $0, \pi, \frac{\pi}{2}, -\frac{\pi}{2}$ correspond respectively to the gates $I, Z, S$ and $S^{\dagger}$, up to global phases). 

The previous definition of FGU and of the corresponding matchgate circuits can be extended to include transformations satisfying Eq.~\eqref{eq:adjoint_action_fgu} with $Q\in\Orth{2n}$, which we refer to as generalized FGU. Notice that most of the previous discussion on FGU remains valid for these transformations, and in particular Eqs.~\eqref{eq:fgu_on_majorana_products} and~\eqref{eq:composition_fgu} hold true. Generalized FGU form a group comprising the previous FGU together with unitary transformations implementing reflections, namely transformations that map $\g_{k}$ to $-\g_{k}$ for some $k\in\left[2n\right]$ and leave all over Majorana operators invariants. Adding such reflections allows to navigate between both connected components of $\Orth{2n}$ corresponding to matrices with determinant $+1$ and $-1$. Note that it suffices to add a single reflection to the group of FGU to generate the whole group of generalized FGU. A simple choice is to include the reflection with respect to the last Majorana operator $\g_{2n}$. In terms of qubits, this amount to allow the addition of the single-qubit Pauli gate $X_{n}$ to the previous matchgate circuits, as can be checked from Eq.~\eqref{eq:Majorana_Pauli_map}. We refer to the corresponding circuits as generalized matchgate circuits. Remark that by using Eq.~\eqref{eq:Majorana_Pauli_map} one may verify that the reflections with respect to $\g_{2k-1}$ and $\g_{2k}$ correspond respectively to $X_{k}\prod_{l>k} Z_{l}$ and $Y_{k}\prod_{l>k} Z_{l}$, which can easily be decomposed into products of $X_{n}$ and matchgates.

Due to the connection between matchgate circuits and FGU, we write respectively $\Mn^{+}$ and $\Mn$ for the groups of FGU and generalized FGU~\footnote{Here the superscript indicates that $\Mn^{+}$ can be seen as the subgroup of $\Mn$ whose elements $U_Q$ are associated to $(+1)$-determinant matrices $Q$. Note that other works such as Ref.~\cite{helsenMatchgateBenchmarkingScalable2022} use the opposite conventions.}.

\subsection{Previous results}
\label{subsec:previous-results}
Having reintroduced the relevant ensembles, we can now recall some of the results on FGU shadows protocols established in previous works. \citet{zhaoFermionicPartialTomography2021} derived an exact expression for the measurement channel for the ensemble $\Mn^{+}\cap\Cliff$ that reads
\bea
\Mcal_{\Mn^{+}\cap\Cliff} = \sum_{k=1}^{n}\binom{n}{k}\binom{2n}{2k}^{-1}\Pcal_{2k}
\label{eq:measurement_channel_fgu}
\eea
with $\Pcal_{2k}$ the projector on the subspace of Majorana operators of degree $2k$. This measurement channel is diagonal in the basis of Majorana operators and we write ${\lambda_{k,n} := \binom{n}{k}\binom{2n}{2k}^{-1}}$ its eigenvalues. Thanks to its diagonal form, this channel can be easily inverted. Moreover, a simple calculation shows that for a $2k$-degree Majorana operator $\g_{\bsmu}$ the shadow norm is given by $\lambda_{k,n}^{-1}$, which yields an upper bound on the variance of the corresponding shadow estimator. The authors prove that this bound is optimal in the sense that there is no subgroup of Clifford FGU resulting in a strictly lower shadow norm~\footnote{Notice that this result does not forbid the existence of a sub-ensemble of FGU that would yield a lower variance for the shadow estimators in general. For instance, one could consider ensembles of FGU being non-Clifford, without a group structure or equipped with a non-uniform distribution.}. Another important result from Ref.~\cite{zhaoFermionicPartialTomography2021} is the fact that the measurement channel of an ensemble of FGU associated with a set of generalized permutation matrices is independent of the signs of the matrices entries~\footnote{This result is essentially contained in Lemma 3 of the Supplementary Material of Ref.~\cite{zhaoFermionicPartialTomography2021}. In particular, it holds even for generalized permutation matrices of cyclic order $d$ whose entries take values in the group of the $d$-th roots of unity.}. In particular, the authors show that the sub-ensembles of Clifford FGU corresponding to the groups $\Sym^{+}(2,2n)$ and $\Sym^{+}(2n)$ result in the same measurement channel given by Eq.~\eqref{eq:measurement_channel_fgu}. This result seems to indicate an equivalence between the different ensembles of Clifford FGU. However, the equality of the measurement channels does not guarantee that the corresponding shadow estimators have the same variance, even though the authors showed that the corresponding shadow norms are the same. Besides, a potential limitation of the approach taken in Ref.~\cite{zhaoFermionicPartialTomography2021} lies in the restriction to the subgroup of Clifford FGU. In fact, focusing on this subgroup appears to single out a preferred basis of Majorana operators, and the bounds on the variance of the corresponding shadow estimators do not necessary hold for rotated bases of Majorana operators of the form $\tilde{\g}_{k} := \Ucal_{Q}(\g_{k})$ for some $Q\in\Orth{2n}$.

Building on the work of Ref.~\cite{zhaoFermionicPartialTomography2021}, \citet{wanMatchgateShadowsFermionic2023} proposed to use ensembles of generalized matchgate circuits. Most of their results stem from an important property which they prove, namely that the $3$-fold channels of the ensembles $\Mn$ and $\Mn\cap\Cliff$ are equal:
\bea
\Ecal^{(3)}_{\Mn} = \Ecal^{(3)}_{\Mn\cap\Cliff}\,.
\label{eq:3-fold-mn-and-mncliff}
\eea
This result is reminiscent of the 3-design property of the Clifford group~\cite{webbCliffordGroupForms2016a,zhuMultiqubitCliffordGroups2017,mitsuhashiCliffordGroupUnitary2023}, and the authors informally summarize it by saying that the group of Clifford generalized matchgate circuits form a matchgate 3-design. As a consequence of this result, the ensembles $\Mn\cap\Cliff$ and $\Mn$ are equivalent for the classical shadows protocol, and they yields the same measurement channels as well as shadow estimators with the same variance. Moreover, it is shown that the measurement channel is equal to the one obtained in Ref.~\cite{zhaoFermionicPartialTomography2021} for $\Mn^{+}$, such that
\bea
\Mcal_{\Mn} = \Mcal_{\Mn\cap\Cliff} = \Mcal_{\Mn^{+}\cap\Cliff}\,.
\eea
The authors also derive improved variance bounds for Majorana operators as well as for other types observables, including some important overlaps quantities for hybrid quantum-classical Monte Carlo simulations. The equivalence between $\Mn\cap\Cliff$ and $\Mn$ enables to use the symmetries of the Clifford group while preserving the invariance under rotations and reflections of the generalized matchgate group. This allows the authors to evade the specification of a preferred set of operators and to apply their bounds to any rotated bases of Majorana operators. 

\citet{ogormanFermionicTomographyLearning2022} extend the results of Ref.~\cite{zhaoFermionicPartialTomography2021} and provide a simplified analysis of the protocol as well as a corrected expression of the variance of the shadow estimators. Moreover, they show that the measurement channel of any sub-ensemble of $\Mn^{+}\cap\Cliff$ only depends on the perfect-matching associated with each (signed) permutation of the ensemble. The perfect-matching associated with a permutation $\sigma:\left[2n\right]\mapsto\left[2n\right]$ is defined as
\bea
\mathrm{PerfMatch}(\sigma):=\left\{\left\{\sigma(2i-1),\sigma(2i)\right\}\,,\;i\in[n]\right\}\,.
\eea
As such, $\mathrm{PerfMatch}(\sigma)$ is the equivalence class of permutations that differs from $\sigma$ by transposition acting on pairs of the form $(2i-1, 2i)$ and by permutations preserving these pairs. For $Q\in\Sym(2,2n)$, we define $\mathrm{PerfMatch}(Q)$ as the perfect-matching of the corresponding unsigned permutation. We denote $\mathrm{PerfMatch}(2n)$ the set of all possible perfect-matching of $\left[2n\right]$. A complete set of representatives of $\mathrm{PerfMatch}(2n)$ is a subset of $\Sym(2,2n)$ composed of exactly one representative per class in $\mathrm{PerfMatch}(2n)$. Using their result, the authors show that any such set equipped with a uniform measure leads to the same measurement channel as $\Mn^{+}\cap\Cliff$. Hence, they find a strict sub-ensemble of $\Mn^{+}\cap\Cliff$ admitting the same measurement-channel. As before, note that this imply an equality of the corresponding shadow norms but not necessarily of the variance of the shadow estimators. Besides their results, the authors conjecture the existence of an efficient sampling scheme for some complete sets of representatives of $\mathrm{PerfMatch}(2n)$.

Other works in the literature investigated the use of FGUs for classical shadows protocols. \citet{wuErrormitigatedFermionicClassical2024} presented an error mitigated version of the protocol of Ref.~\cite{zhaoFermionicPartialTomography2021}. Statistical properties of the matchgate shadows of Ref.~\cite{wanMatchgateShadowsFermionic2023} were investigated in Refs.~\cite{scheurerTailoredExternallyCorrected2024}~and~\cite{kiserClassicalQuantumCost2024} in the respective context of quantum chemistry and QC-QMC. This protocol was also numerically and experimentally investigated in Ref.~\cite{huangEvaluatingQuantumclassicalQuantum2024}, again in view of QC-QMC simulations. The authors show that the matchgate shadows protocol is robust to noise, connecting with earlier results of this nature~\cite{chenRobustShadowEstimation2021, kohClassicalShadowsNoise2022}. They also show that the post-processing of the data remains a challenging bottleneck for an application to QC-QMC. Another classical shadows scheme based on the subset of number-conserving generalized FGU was investigated in Ref.~\cite{lowClassicalShadowsFermions2022} for the particular case of quantum states with a fixed particle number. \citet{lowClassicalShadowsFermions2022} found an exponential improvement for average variance of their estimator over the worst-case bounds obtained in Ref.~\cite{zhaoFermionicPartialTomography2021}. However, this protocol suffers some limitations that are discussed in Ref.~\cite{wanMatchgateShadowsFermionic2023}. \citet{zhaoGrouptheoreticErrorMitigation2023} propose a quantum error-mitigation strategy using classical shadow tomography and relying on symmetries of the system of interest. They apply their method in the context of matchgate shadows protocols and derive an optimal sampling of the continuous matchgate group. At last, the generalized FGU group was also explored in Ref.~\cite{helsenMatchgateBenchmarkingScalable2022} in the context of randomized benchmarking.

The different ensembles of generalized matchgates studied in this work and their corresponding matrix groups are summarized in Table~\ref{tab:matchgate_ensembles}.
\begin{table}[t]
\centering
\begin{tabular}{c | c | c}
    $\mathbb{U} := \left\{\,\Ucal_{Q},\, Q\in \mathrm{S}\,\right\}$ & $\mathrm{S}\subset \Orth{2n}$ & Ref. \\
    \hline
    $\mathrm{M}_n$ & $\Orth{2n}$ & \cite{wanMatchgateShadowsFermionic2023} \\
    $\mathrm{M}_n\cap \Cliff$ & $\mathrm{Sym}(2,2n)$ & \cite{wanMatchgateShadowsFermionic2023}\\
    $\mathrm{M}^{+}_n$ & $\SO{2n}$ & This work \\
    $\mathrm{M}^{+}_n\cap \Cliff$ & $\mathrm{Sym}^{+}(2,2n)$ & \cite{zhaoFermionicPartialTomography2021} \\
    n.a. & $\mathrm{Sym}^{+}(2n)$ & \cite{zhaoFermionicPartialTomography2021}\\
    n.a. & $S\cong\mathrm{PerfMatch}(2n)$ & \cite{ogormanFermionicTomographyLearning2022}\\
\end{tabular}
\caption{Sub-ensembles of the generalized matchgate group used in the literature and the corresponding subsets ${S\subseteq\Orth{2n}}$. The last line correspond to subsets that are complete sets of representatives of $\mathrm{PerfMatch}(2n)$, i.e. sets that are isomorphic to $\mathrm{PerfMatch}(2n)$ under the associated natural projection.}
\label{tab:matchgate_ensembles}
\end{table}

\section{Locally Random Matchgate Ensembles and Clifford 3-cubatures}
\label{sec:random_matchgate_circuits_and_Clifford-cubatures}

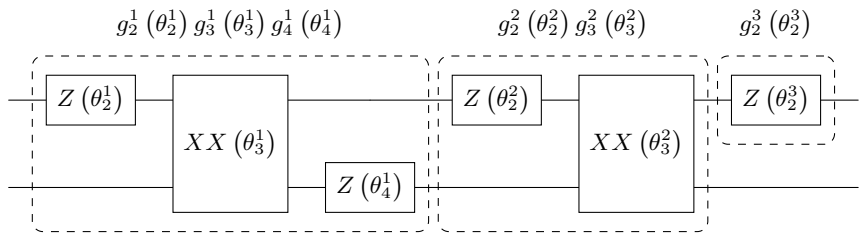
\begin{figure*}
    \begin{quantikz}[thin lines] 
        & \gate{Z\left(\theta^{1}_{2}\right)}\gategroup[2,steps=3,style={dashed,rounded corners, inner xsep=2pt}, label style={yshift=0.2cm}]{$g^{1}_{2}\left(\theta^{1}_{2}\right)g^{1}_{3}\left(\theta^{1}_{3}\right)g^{1}_{4}\left(\theta^{1}_{4}\right)$} & \gate[2]{{XX}\left(\theta^{1}_{3}\right)} & & \gate{Z\left(\theta^{2}_{2}\right)}\gategroup[2,steps=2,style={dashed,rounded corners, inner xsep=2pt}, label style={yshift=0.2cm}]{$g^{2}_{2}\left(\theta^{2}_{2}\right)g^{2}_{3}\left(\theta^{2}_{3}\right)$} & \gate[2]{{XX}\left(\theta^{2}_{3}\right)} & \gate{Z\left(\theta^{3}_{2}\right)}\gategroup[1,steps=1,style={dashed,rounded corners, inner xsep=2pt}, label style={yshift=0.2cm}]{$g^{3}_{2}\left(\theta^{3}_{2}\right)$} & \\
        & & & \gate{Z\left(\theta^{1}_{4}\right)} & & & &
    \end{quantikz}
    \caption{Example of matchgate circuit corresponding to the decomposition of Eq.~\eqref{eq:decomposition_in_rotations} for the Jordan-Wigner mapping and $n=2$ qubits.}
    \label{fig:matchgate_circuit_template_n=2}
\end{figure*}

In this section we investigate the use of the ensemble $\Mn^{+}$ corresponding to the matrix group $\SO{2n}$ for the classical shadows protocol. As pointed out in Ref.~\cite{wanMatchgateShadowsFermionic2023}, so far it was unclear whether the matchgate 3-design property holds for $\Mn^{+}$. Here, we provide a positive answer to this question and show that a similar property holds for a larger class of random quantum circuits. Consequently, we obtain that the group $\Mn^{+}$ leads to a classical shadows protocol that is equivalent to one of the group $\Mn^{+}\cap\Cliff$ analyzed in Ref.~\cite{zhaoFermionicPartialTomography2021}.

Our results rely on a well-known decomposition of elements of $\SO{2n}$ into Givens rotations as well as an associated sampling method introduced in Ref.~\cite{hurwitzUberErzeugungInvarianten1897}. Consider a matrix ${Q\in\mathrm{SO}(2n)}$, one can show that there exists Givens rotations $g^{l}_{k}(\theta^{l}_{k})$ acting on axes ${(k-1,k)}$ with ${1\leq l < 2n+2-k \leq 2n}$ such that
\bea
    Q = (g^{1}_{2}g^{1}_{3}\dots g^{1}_{2n})\dots(g^{2n-2}_{2}g^{2n-2}_{3})(g^{2n-1}_{2})\,,
\label{eq:decomposition_in_rotations}
\eea
where we dropped the angles for clarity. Detailed proofs of this decomposition can be found in Refs.~\cite{meckesRandomMatrixTheory2019, diaconisHurwitzOriginsRandom2016}. To this decomposition corresponds a matchgate circuit. Figure~\ref{fig:matchgate_circuit_template_n=2} show this circuit for $2n=2$. In the general case, that circuit is composed of $n(2n-1)$ rotations of angles ${\theta^{l}_{k}}$ that can be adjusted to generate any FGU.  By sampling the rotation angles according to the right distributions, one can use this circuit to sample uniformly over $\Mn^{+}$. This result is encapsulated in the following proposition, which is adapted of a result given in Ref.~\cite{meckesRandomMatrixTheory2019}. Details on our modifications of the original proposition can be found in Appendix~\ref{app:euler_angles_distrib_adaptation}.

\begin{prop}[Proposition 1.6 in Ref.~\cite{meckesRandomMatrixTheory2019}, adapted.]{Let $Q$ be a random matrix defined by Eq.~\eqref{eq:decomposition_in_rotations} for random independent angles $\theta^{l}_{k}$. If
\bea
\theta^{l}_{k}\sim f_{i}(\theta) := \frac{\Gamma\left(\frac{k}{2}\right)}{2\Gamma\left(\frac{1}{2}\right)\Gamma\left(\frac{k-1}{2}\right)}\abs{\sin(\theta)}^{k-2}
\label{eq:haar_distribution_angle_i}
\eea
for all $l$ and $k$, then $Q$ is uniformly distributed on $\SO{2n}$.}
\label{prop:decomposition_uniform_matchgate_group}
\end{prop}
Note that there exists other decompositions similar to the one presented here, some of which rely on Givens rotations with non-adjacent axes (see for instance Ref.~\cite{heissDistributionsAnglesRandom1994}), and other ones with a different order of the Givens rotations (see Ref.~\cite{diaconisBoundsKacMaster2000}) .

To present our result, it is convenient to first introduce a type of ensemble that generalizes the notion of $t$-design to non-uniform distributions, which we call $t$-cubatures. Akin to unitary designs that were introduced as the unitary analogs of spherical designs, we define unitary $t$-cubatures as the analogs of positive cubatures appearing in the literature on numerical integration (see for instance Refs.~\cite{sobolevTheoryCubatureFormulas2006, coolsConstructingCubatureFormulae1997, delaharpeCubatureFormulasGeometrical2005a, goethalsCubatureFormulaePolytopes1981, prestinCubatureFormulasSphere2006}).  Recall that the $t$-fold channel  $\Ecal_{\mathbb{U}}^{(t)}$ of a unitary ensemble $\mathbb{U}$ is given by Eq.~\eqref{eq:t-fold-channel}.

\begin{definition}[Unitary $t$-cubature]
Let $\mathbb{U}\subseteq \mathrm{U}(\Hcal_{n})$ be a unitary ensemble. We say that a finite sub-ensemble $\left\{ U_{j},\, j\in J\right\}\subseteq \mathbb{U}$ with an associated probability distribution $\left(p_j\right)_{j\in J}$ is a $t$-cubature of $\mathbb{U}$ (or $(\mathbb{U}, t)$-cubature) if it satisfies
\bea
\Ecal^{(t)}_{\mathbb{U}} = \sum_{j\in J} p_j \Ucal^{\otimes t}_{j}\,. 
\eea
We call a Clifford $t$-cubature any $t$-cubature whose elements belong to $\Cliff$, and we say that $\mathbb{U}$ admits a (Clifford) $t$-cubature if there exists a (Clifford) $(\mathbb{U},t)$-cubature.
\label{def:t-cubature}
\end{definition}
Consistently with the definition of unitary $t$-design, we will call a $(\mathbb{U},t)$-design any $(\mathbb{U},t)$-cubature associated with a uniform distribution $p_j = 1/\abs{J}, \forall j \in J$. Equipped with the previous definition, we can now state the main result of this section.

\begin{thm}
Let $\mathbb{U}$ be a unitary ensemble generated by a quantum circuit composed of fixed Clifford gates and Pauli rotations with independent random angles distributed symmetrically about the Clifford angles. Then $\mathbb{U}$ admits a Clifford $3$-cubature.
\label{thm:clifford-cubatures}
\end{thm}

Recall that a random angle $\theta$ is symmetrically distributed about an angle $\theta_{0}$ if and only if ${\expect{f(\theta-\theta_0)}=\expect{f(\theta_{0}-\theta)}}$ for any bounded function $f$. To prove our theorem, we only need to focus on unitary ensembles generated by a random Pauli rotation of the form
\bea
\left\{U(\theta) := \exp\left(-i\frac{\theta}{2}P\right),\; \theta\sim \nu\right\}
\label{eq:random_pauli_rotation_ensemble}
\eea
with $P\in\Pau$ and $\nu$ a probability measure on $(-\pi,\pi]$. In fact, for an ensemble $\mathbb{U}$ generated by a circuit with a fixed architecture composed of Clifford gates and independent random rotations, if the unitary ensembles corresponding to the random rotations admit a Clifford $t$-cubature, then so does $\mathbb{U}$ by linearity. Furthermore, as for every Pauli string $P$ there exists a Clifford unitary $C\in\Cliff$ such that $C^{\dagger}PC = Z_{1}$, one can simply focus on the particular case of a single-qubit random $Z$-rotation. For instance, Fig.~\ref{fig:decomposition_xx_in_cliff_and_z} gives such a decomposition for the rotations generated by $P=X\otimes X$ which represents Givens rotations under the JW mapping. Let us denote ${R}_{\theta} := e^{-i\frac{\theta}{2}Z}$, and recall that up to a global phase we have
\bea
{R}_{0} = I,\;{R}_{\frac{\pi}{2}} = S,\;{R}_{\pi} = Z,\; {R}_{\frac{3\pi}{2}} = S^{\dagger}\,.
\eea
As before, we write respectively $\Rcal_{\theta},\Ical, \Scal, \Zcal$ and $ \Scal^{\dagger}$ the corresponding unitary channels. We prove the following lemma in Appendix~\ref{app:proof_lemma_z_rotations}.

\begin{restatable}{lem}{rotationlemma}
{Let $\nu$ be a probability distribution on $(-\pi,\pi]$ symmetric about the Clifford angles and $\theta\sim\nu$. Then
\bea
\expect{\Rcal_{\theta}^{\otimes 3}} = \frac{(1-p)}{2}\left(\Ical^{\otimes 3}+\Zcal^{\otimes 3}\right)+ \frac{p}{2}\left(\Scal^{\otimes 3}+\Scal^{\dagger\otimes 3}\right)
\label{eq:rotation-3-fold-channel}
\eea
with $p:=\expect{\sin(\theta)^{2}}$.
}
\label{lemma:3-fold_z_rotation}
\end{restatable}
This shows that under the constraints on angle distribution state in the Lemma, the unitary ensemble corresponding to a random $Z$-rotation admits a Clifford $3$-cubature, and the proof of Theorem~\ref{thm:clifford-cubatures} follows.

Note that Theorem~\ref{thm:clifford-cubatures} and Lemma~\ref{lemma:3-fold_z_rotation} above are generalizations of previous results presented in Ref.~\cite{heyraudEfficientEstimationTrainability2023a} that focused on Clifford $2$-cubatures. Intuitively, these results can be interpreted as the outcomes of a decoherence effect induced by a random choice of the angles. It is natural to wonder whether or not the previous lemma generalizes to $k$-fold channels with $k\geq 4$. We provide a negative answer to this question in the following lemma, which proof is deferred to Appendix~\ref{app:no-Clifford-4-cubature}.

\begin{restatable}{lem}{nofourfoldlemma}
{Let $\nu$ be a probability distribution on $(-\pi,\pi]$ symmetric about the Clifford angles and $\theta\sim\nu$. If $\nu$ is not supported by the set of Clifford angles, then the ensemble $
\left\{\Rcal_{\theta}, \theta\sim\nu\right\}
$ admits no Clifford $4$-cubature.
}
\label{lemma:no-go-4-fold}
\end{restatable}
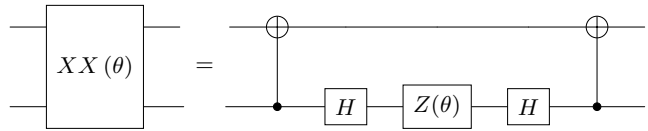
\begin{figure}[t]
\begin{adjustbox}{width=\columnwidth}
\begin{quantikz}[thin lines] 
        &  \gate[2]{{XX}\left(\theta\right)} & \midstick[2,brackets=none]{=} & \targ & & & & & \targ & & \\
& & & \ctrl{-1} & \gate{H} & \gate{Z(\theta)} & \gate{H} & \ctrl{-1} &
\end{quantikz}
\end{adjustbox}
\caption{Decomposition of the $2$-qubits rotation generated by $X\otimes X$ into Clifford gates and single-qubit $Z$-rotation.}
\label{fig:decomposition_xx_in_cliff_and_z}
\end{figure}
For a single qubit, this result shows that the matchgate group does not admit any 4-Clifford cubature. However this does not necessary imply that this is the case for $n\geq 2$, which remains an open question for future work. The $t$-fold channels of unitary ensembles admitting a Clifford $t$-cubature are stabilizer-preserving channels, for which efficient classical simulation schemes exists (see for instance Ref.~\cite{howardApplicationResourceTheory2017a}). Remark that determining the degree of non-stabilizerness of a quantum state or operation (often referred to as the ``magic'' in the literature) is a delicate task~\cite{seddonQuantifyingMagicMultiqubit2019a, haugScalableMeasuresMagic2023}. Interestingly, the previous lemma suggests that correlations between random single-qubit Pauli rotations can be a source of magic for the corresponding $1$-fold average channel.

Theorem~\ref{thm:clifford-cubatures} covers matchgate circuits with a fixed architecture and whose random angles satisfy the required symmetry constraints. In particular, as the distributions of Eq.~\eqref{eq:haar_distribution_angle_i} are symmetric with respect to the Clifford angles, the theorem applies to the random matchgate circuits corresponding to the decomposition of Proposition~\ref{prop:decomposition_uniform_matchgate_group}. Note that this is independent of the exact fermion-to-qubit mapping, as long as Majorana operators are mapped to Pauli strings. This allows us to prove the following proposition, which generalizes the matchgate $3$-design result of Ref.~\cite{wanMatchgateShadowsFermionic2023} to $\Mn^{+}\cap\Cliff$.

\begin{prop}
The uniform FGU group $\Mn^{+}$ admits a Clifford $3$-design, i.e.
\bea
\Ecal^{(3)}_{\Mn^{+}} = \Ecal^{(3)}_{\Mn^{+}\cap\Cliff}\,.
\eea
\label{prop:matchgate_3-fold}
\end{prop}
From Theorem~\ref{thm:clifford-cubatures}, there exists a Clifford $3$-cubature ${\{\Ccal_{j},\;j\in J\}\subset \Mn^{+}\cap\Cliff}$ with a probability distribution $(p_j)_{j\in J}$ such that
\bea
\Ecal^{(3)}_{\Mn^{+}} = \sum_{j\in J}p_{j}\Ccal_{j}^{\otimes 3}\,.
\label{eq:3-cubature-matchgate-proof}
\eea
To prove the proposition, one only needs to show that ${(p_j)_{j\in J}}$ is the uniform distribution on ${\Mn^{+}\cap\Cliff}$. Remark that the channels in $\Mn^{+}\cap\Cliff$ are linearly independent in $\Lcal(\Lcal(\Hcal_{n}))$, so that the decomposition of Eq.~\eqref{eq:3-cubature-matchgate-proof} is unique. This can be verified by using the Pauli transfer matrix representations, that is the Liouville matrix representation of the channels in the basis of Pauli operators (see e.g. Ref.~\cite{woodTensorNetworksGraphical2015}). The right invariance of the Haar measure on $\Mn^{+}$ gives $\forall k\in J$ and $\forall \Ccal\in\Mn^{+}\cap\Cliff$:
\bea
\Ecal^{(3)}_{\Mn^{+}} &= \Ecal^{(3)}_{\Mn^{+}}\left(\Ccal_{k}^{\dagger}\Ccal\right)^{\otimes 3}\\
&= \sum_{j\in J}p_{j}\left(\Ccal_{j}\Ccal_{k}^{\dagger}\Ccal\right)^{\otimes 3}\,.
\eea
Hence, by uniqueness of the decomposition we have ${\{\Ccal_{j},\, j\in J\} = \Mn^{+}\cap\Cliff}$ and taking $\Ccal = \Ccal_{l}$ we get ${p_{k}=p_{l}},\,\forall k,l\in J$. As a result
\bea
\sum_{j\in J} p_j \Ccal_{j}^{\otimes 3} = \Ecal^{(3)}_{\Mn^{+}\cap\Cliff}\,,
\eea
which achieves the proof.

The results of this section can be summarized as follow. First, we provided a decomposition of the elements of the uniform FGU ensemble $\Mn^{+}$ associated with the matrix group $\SO{2n}$ into products of independent Givens rotations through Eq.~\eqref{eq:decomposition_in_rotations} and Proposition~\ref{prop:decomposition_uniform_matchgate_group}. Under the considered fermion-to-qubit mapping, this yielded a decomposition of the corresponding matchgates circuits in terms of independent random Pauli rotations. Second, we proved in Lemma~\ref{lemma:3-fold_z_rotation} that up to the third order such random Pauli rotations can be written as convex sums of Clifford gates, provided their random rotation angle is symmetrically distributed according to the Clifford angles. This allowed us to prove the existence of a Clifford 3-cubature for a large class of random matchgates circuits in Theorem~\ref{thm:clifford-cubatures}. Leveraging this theorem and the previous decomposition, we obtained the existence of a Clifford 3-cubature for $\Mn^{+}$. Finally, we used the invariance of the Haar measure on this group to derive the equality of the $3$-fold channels of $\Mn^{+}$ and of its Clifford subgroup $\Mn^{+}\cap\Cliff$, thereby proving the equivalence of the shadows protocols associated with the matrix groups $\SO{2n}$ and $\Sym^{+}(2,2n)$.

\section{Invariances and equivalence of the Matchgate Shadows Protocols}
\label{sec:invariances_and_equivalences}

The results of the previous section proves that the classical shadows protocols corresponding to the ensembles $\Mn^{+}$ and $\Mn^{+}\cap\Cliff$ are equivalent. In this section we extend the results of Refs.~\cite{zhaoFermionicPartialTomography2021} and~\cite{ogormanFermionicTomographyLearning2022} and show that, as for the measurement channels, the variances of the shadow estimators associated with the ensembles of generalized Clifford FGU corresponding to signed permutations are independent of the permutations signs and only depend on the associated perfect-matchings. From this, we identify the relevant properties of ensembles of matchgate circuits for the classical shadows protocol and we establish the equivalence between the different ensembles of Table~\ref{tab:matchgate_ensembles}. Importantly, this result allows us to transfer the performances guarantees associated with a given ensemble to the others. For instance, the results obtained in Ref.~\cite{wanMatchgateShadowsFermionic2023} for $\Mn$  and $\Mn\cap\Cliff$, which rely on the invariance of the shadows protocols under an arbitrary rotation of the basis of Majorana operators, can be generalized to the other ensembles.

One of the key ingredient used by \citet{wanMatchgateShadowsFermionic2023} to derive their results is the invariance of the generalised FGU group $\Mn$ under arbitrary reflections. This invariance enables them to explicitly calculate the $3$-fold channel $\Ecal^{(3)}_{\Mn}$ and to establish the equality ${\Ecal^{(3)}_{\Mn} = \Ecal^{(3)}_{\Mn\cap\Cliff}}$. Before we state and prove our results, let us introduce a few facts regarding reflections. These transformations are intimately related with the signs of the matrices in $\Sym(2,2n)$ associated with Clifford generalized FGU. In fact, the channels implementing reflections with respect to Majorana operators are of the form $\Ucal_{D}$ with $D\in\mathbb{Z}^{2n}_{2}$. Consider the channel $\Ucal_{Q}$ with $Q\in\Sym(2,2n)$. Recall that $Q$ admits a unique decomposition $Q = DP$ with $D\in\mathbb{Z}^{2n}_{2}$ and $P\in\Sym(2n)$, such that $\Ucal_{Q} = \Ucal_{P}\Ucal_{D}$. For elements of $\Sym(2,2n)$, define the equivalence relation $Q\sim Q'$ if and only if the entries of $Q$ and $Q'$ only differ by a sign, i.e. $Q=DQ'$ for some $D\in\mathbb{Z}^{2n}_{2}$. This equivalence relation can be lifted to the corresponding quantum channel and we write $\Ucal_{Q}\sim \Ucal_{Q'}$ if and only if $Q\sim Q'$. In the following, we will prove that for a unitary ensemble $\mathbb{U}\subset \Mn\cap\Cliff$, the corresponding classical shadows protocol only depends on the equivalence classes of the elements of $\mathbb{U}$ for the equivalence relation $\sim$. 

We define the $k$-fold channels associated with the ensemble of reflections (equipped with the corresponding Haar measure) as
\bea
    \Lambda^{(k)} &:= \frac{1}{2^{2n}}\sum_{D\in \mathbb{Z}^{2n}_{2}} \Ucal^{\otimes k}_{D}\,.
\eea
By the right invariance of the Haar measure, we have $\Ucal^{\otimes k}_{D}\Lambda^{(k)} =\Lambda^{(k)}$ for all $D\in\mathbb{Z}_{2}^{2n}$. In particular, the following lemma holds true.
\begin{lem}
    Let $\Ucal_{Q},\, \Ucal_{Q'}$ be a Clifford generalized FGUs such that $\Ucal_{Q}\sim \Ucal_{Q'}$. Then we have $\forall k\in\mathbb{N}^{\ast}$:
    \bea
        \Ucal^{\otimes k}_{Q}\Lambda^{(k)} = 	\Ucal^{\otimes k}_{Q'}\Lambda^{(k)}\,.
        \label{eq:equivalence_sign_k-fold}
    \eea
    \label{lem:equivalence_sign_k-fold}
\end{lem}	
We can now prove the following proposition.

\begin{prop}
    Let $\mathbb{U}\subseteq \Mn\cap\Cliff$ be a unitary ensemble of Clifford generalized FGU. For any $\Ucal_{Q}\in\mathbb{U}$ with $Q\in\Sym(2,2n)$, replacing $\Ucal_{Q}$ by some $\Ucal_{Q'}\sim\Ucal_{Q}$ has no effect on the measurement channel and on the variance of the estimators of the resulting classical shadows protocol.
    \label{prop:invariance_classical_shadows_reflections}
\end{prop}

The invariance of measurement channel under such exchange was proven in Ref.~\cite{zhaoFermionicPartialTomography2021}. We can thus focus on the variance of the shadow estimators. Let us write ${\{\Ucal_{Q_{j}},\, j\in J\}}$ the elements of $\mathbb{U}$ and $(p_j)_{j\in J}$ the associated probability distribution. Denote $\Ecal_{\mathbb{U}}^{(3)} = \sum_{j\in J} p_j\Ucal^{\otimes 3}_{Q_j}$ the corresponding $3$-fold channel. As recalled in Sec.~\ref{subsec:classical_shadows}, for an observable $O\in\Lcal(\Hcal_{n})$, the variance of the corresponding estimator $\hat{o}$ is essentially determined by the second raw moment $\expect{\hat{o}^{2}}$ which expression is given in Eq.~\eqref{eq:second-raw-moment}. Using the linearity of the trace, we can rewrite this expression
\bea
\expect{\hat{o}^{2}} = \tr{\Upsilon_{\mathbb{U}}^{(3)}(\rho\otimes\Mcal^{-1}(O_i)\otimes\Mcal^{-1}(O_i))}
\eea
where we defined
\bea
\Upsilon_{\mathbb{U}}^{(3)} := \Ecal_{\mathbb{U}}^{(3)}\left(\sum_{z\in\{0,1\}^{n}}\dyad{z}^{\otimes 3}\right)\,.
\label{eq:upsilon}
\eea
Remark that $\forall P\in \Pau$ the map $A\mapsto P^{\dagger}AP$ is bijective and maps bitstring states of the form $\dyad{z}$ to bitstring states, such that
\bea
P^{\dagger\otimes 3}\left(\sum_{z\in\{0,1\}^{n}}\dyad{z}^{\otimes 3}\right)P^{\otimes 3} = \sum_{z\in\{0,1\}^{n}}\dyad{z}^{\otimes 3}\,.
\eea
As reflections with respect to Majorana operators corresponds to Pauli strings under the JW mapping, we have
\bea
\Lambda^{(3)}\left(\sum_{z\in\{0,1\}^{n}}\dyad{z}^{\otimes 3}\right) =\sum_{z\in\{0,1\}^{n}}\dyad{z}^{\otimes 3}\,
\eea
and one can rewrite Eq.~\eqref{eq:upsilon} as 
\bea
\Upsilon^{(3)}_{\mathbb{U}} = \Ecal^{(3)}_{\mathbb{U}}\Lambda^{(3)}\left(\sum_{z\in\{0,1\}^{n}}\dyad{z}^{\otimes 3}\right)\,.
\eea
Hence we can replace $\Ecal_{\mathbb{U}}^{(3)}$ by $\Ecal_{\mathbb{U}}^{(3)}\Lambda^{(3)}$ in the expression of the variance. Finally, one can rewrite
\bea
\Ecal_{\mathbb{U}}^{(3)}\Lambda^{(3)} = \sum_{j\in J}p_j\Ucal^{\otimes 3}_{Q_i}\Lambda^{(3)}
\eea
and invoking Lemma~\ref{lem:equivalence_sign_k-fold} achieves the proof. Note that adapting this proof to the measurement channel is straightforward, so that we incidentally proved the result of Ref.~\cite{zhaoFermionicPartialTomography2021}.

In the cases where elements of the considered ensemble $\mathbb{U}$ are obtained as products of more elementary Clifford generalized FGUs, we can likewise replace the elementary channels by any other equivalent channel. This is a direct consequence of Proposition~\ref{prop:invariance_classical_shadows_reflections} and of the following Lemma:
\begin{lem}
    Let $\Ucal_{Q}$ be a Clifford generalized FGU, then
    \bea
    \Ucal^{\otimes k}_{Q}\Lambda^{(k)} = \Lambda^{(k)}\Ucal^{\otimes k}_{Q}\,,\quad \forall k\in\mathbb{N}^{\ast}
    \eea
    \label{lem:commutation_reflections}
\end{lem}
This lemma is easily proven. To lighten notations we take ${k=1}$, the proof remaining valid for any $k$. Decomposing $Q$ as  $Q = DP$ with $P\in\Sym(2n)$ and $D\in \mathbb{Z}^{2n}_{2}$, we have $\forall D'\in \mathbb{Z}^{2n}_{2}$:
\bea
\Ucal_{Q}\Ucal_{D'} &= \Ucal_{D'Q} = \Ucal_{D'DP}\\
&= \Ucal_{DD'P}\\
&= \Ucal_{DPP^{-1}D'P} \\
&= \Ucal_{P^{-1}D'P}\Ucal_{Q}
\eea
Hence
\bea
\Ucal_{Q}\Lambda^{(1)} &= \frac{1}{2^n}\sum_{D\in\mathbb{Z}_{2}^{2n}}\Ucal_{Q}\Ucal_{D}\\
&=  \left(\frac{1}{2^n}\sum_{D\in\mathbb{Z}_{2}^{2n}}\Ucal_{P^{-1}DP}\right)\Ucal_{Q}\\
&= \Lambda^{(1)}\Ucal_{Q}\,,
\label{eq:commutation_cliff_reflection_1fold}
\eea
using the fact that for any $P\in\Sym(2n)$ the map ${D\mapsto P^{-1}DP}$ is bijective an preserves $\mathbb{Z}^{2n}_{2}$. 

Proposition~\ref{prop:invariance_classical_shadows_reflections} and Lemma~\ref{lem:commutation_reflections} clearly prove that the first five ensembles of Table~\ref{tab:matchgate_ensembles} lead to equivalent classical shadows protocols. More generally, this equivalence holds between any unitary sub-ensembles of the generalized FGU admitting Clifford $3$-cubatures whose elements differ only by reflections. In terms of matchgate circuits, one can rephrase this equivalence and state that inserting Pauli strings into any matchgates circuit of a sub-ensemble of $\Mn\cap\Cliff$ has no effect on the corresponding classical shadows protocol. Note that the addition of reflections plays a crucial in the proof of the results of Ref.~\cite{wanMatchgateShadowsFermionic2023}. From our result, it appears that these reflections naturally stems from the measurement and averaging process of the shadows protocol, so that there is no need to explicitly add them to the circuits of the considered ensemble.

Having proved the previous equivalences, it remains to show that the shadows protocol only depends on the perfect-matching corresponding to permutation matrices of the considered ensemble. \citet{ogormanFermionicTomographyLearning2022} proved this result for the measurement channels using arguments similar to the ones we use in the proof of Proposition~\ref{prop:invariance_classical_shadows_reflections}. As before, we generalize it to the variance of the corresponding estimators.

In what precedes we exploited the symmetry of the state 
$\sum_{z}\dyad{z}^{\otimes 3}$ under conjugation by $3$-fold products of Pauli strings. This state is also clearly invariant under conjugation by any $3$-fold product of single-qubit $Z$-rotations and $CNOT$ gates. Since the transposition $T_i$ that exchange the pair $(\g_{2i-1}, \g_{2i})$ corresponds to the $Z$-rotation on qubit $i$ (up to irrelevant signs), we get that 
\bea
\Ucal_{T_i}^{\otimes}\left(\sum_{z}\dyad{z}^{\otimes 3}\right) = \sum_{z}\dyad{z}^{\otimes 3}\,.
\eea
On the other hand, any permutation $\tilde{Q}$ preserving the pairs of the form $(2i-1, 2i)$ can be represented (again up to irrelevant signs) by products of $SWAP$ gates, which are themselves products of $CNOT$ gates. Hence, for any such generalized permutation, we also have 
\bea
\Ucal_{\tilde{Q}}^{\otimes}\left(\sum_{z}\dyad{z}^{\otimes 3}\right) = \sum_{z}\dyad{z}^{\otimes 3}\,.
\eea
With these invariances, the proof of Proposition~\ref{prop:invariance_classical_shadows_reflections} is straightforwardly adapted and the following proposition holds.
\begin{prop}
    Let $\mathbb{U}\subseteq \Mn\cap\Cliff$ be a unitary ensemble of Clifford generalized FGU. For any $\Ucal_{Q}\in\mathbb{U}$ with ${Q\in\Sym(2,2n)}$, replacing $\Ucal_{Q}$ by some $\Ucal_{Q'}$ with $Q'$ satisfying
    \bea
        \mathrm{PerfMatch}(Q) = \mathrm{PerfMatch}(Q')
    \eea
    has no effect on the measurement channel and on the variance of the estimators of the resulting classical shadows protocol.
    \label{prop:invariance_perf_match}
\end{prop}
This last result achieves the proof that the shadows protocols corresponding to the ensembles of Table~\ref{tab:matchgate_ensembles} are all equivalent. 

\begin{figure*}[ht]
\includegraphics[width=\textwidth]{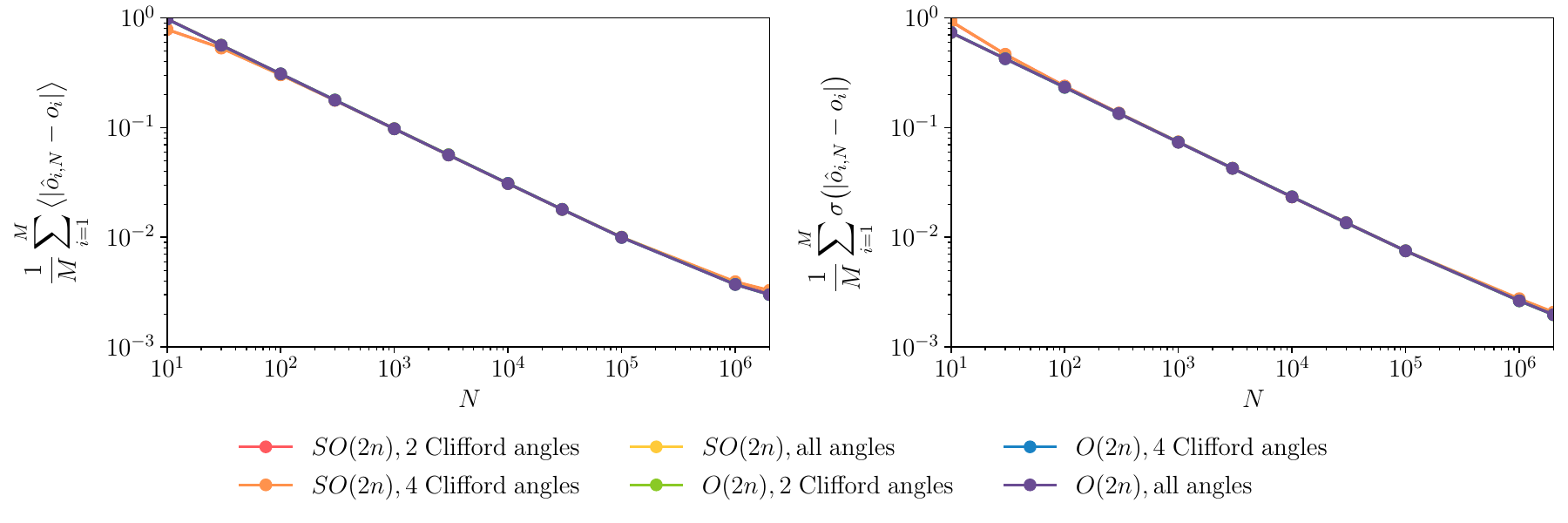}
    \caption{Statistics of the error of the shadow estimators as a function of the number of shadows for the different samplings of the rotation angles. The input state is an 8-qubit state obtained from CCSD calculation on a fictitious $H^{4}$ molecule. The set of observables considered are the Majorana operators of order 2, namely $\left\{\g_p\g_q,\;p<q\,,\;p,q\in\left[2n\right]\right\}$.
    The left panel shows the mean value of the shadow estimators absolute error for a given number $N$ of shadow shots, averaged over all of the observables. The right panel shows the standard deviation of the same quantity. We simulate a total of 2e6 shadows samples, and the results are obtained using a bootstrap sampling with a bootstrap sample size of 1e3.}
    \label{fig:numerical-results}
\end{figure*}

\section{Samplings of ensembles of matchgate circuits}
\label{sec:sampling-schemes}

In this section, we build on the results of Secs.~\ref{sec:random_matchgate_circuits_and_Clifford-cubatures} and~\ref{sec:invariances_and_equivalences} to improve over the existing methods and derive a sampling scheme for a small sub-ensemble of $\Mn^{+}\cap\Cliff$ that is optimal in terms of number of gates and that inherits the performance guarantees of the protocols considered previously. Before we present our optimal sampling scheme, we also introduce simple sampling schemes for sub-ensembles of generalized FGUs resulting in equivalent shadows protocols and based on the results of Section~\ref{sec:random_matchgate_circuits_and_Clifford-cubatures}. Note that a sampling scheme for a sub-ensemble of $\Mn^{+}$ can easily be extended to a sub-ensemble of $\Mn$ by randomly adding a single reflection (for instance with a Pauli $X$ gate on the last qubit) to the generated circuits with probability $1/2$. Consequently, we focus on ensembles in $\Mn^{+}$ in this section.

\begin{figure}
\begin{algorithm}[H]
\raggedright
\caption{Sampling of matchgates circuits for the classical shadows protocol}
\label{algo:sampling}
\vspace{4pt}
    \textbf{Input:} Number of qubits $n$
    
    \textbf{Output:} Random matchgates circuit

    \begin{enumerate}
        \itemsep -0.8\parsep
        \item Sample a permutation $P\in \Sym(2n)$
        \item Extract the corresponding perfect-matching $\mathrm{PerfMatch}(P) = \{\{P_{2i-1}, P_{2i}\}\,,\, i \in [n]\}$
        \item $\forall i\in [n]$, sort the pair $\{P_{2i-1}, P_{2i}\}$ in increasing order
        \item Sort the $n$ pairs by their first element in increasing order
        \item Concatenate the sorted pairs to obtain a new permutation $P'$
        \item Decompose the permutation $P'$ in a sequence of transpositions using the bubblesort algorithm
        \item Compile the quantum circuit by turning each transposition into a Givens rotation with angles $\pi/2$
    \end{enumerate}

\end{algorithm}
\end{figure}

The decomposition of Proposition~\ref{prop:decomposition_uniform_matchgate_group} directly provides a method to sample circuits uniformly in $\Mn^{+}$, relying on single-and-two qubits random rotations. Although simple, this sampling scheme is not necessarily efficient. An efficient sampling schemes for $\Mn^{+}$ can be found in Ref.~\cite{zhaoGrouptheoreticErrorMitigation2023}.

Proposition~\ref{prop:decomposition_uniform_matchgate_group} and Lemma~\ref{lemma:3-fold_z_rotation} yields a simple sampling scheme for $\Mn^{+}\cap\Cliff$. In fact, assuming that the distribution of random angles $\theta$ satisfy the constraints of Lemma~\ref{lemma:3-fold_z_rotation}, Eq.~\eqref{eq:rotation-3-fold-channel} can be rewritten for random Givens rotations $\Gcal_{k}(\theta)$ as:
\bea
\expect{\Gcal^{\otimes 3}_{k}(\theta)} =& \frac{(1-p)}{2}\left(\Gcal^{\otimes 3}_{k}(0)+\Gcal^{\otimes 3}_{k}(\pi)\right)\\ &
+\frac{p}{2}\left(\Gcal^{\otimes 3}_{k}\left(\frac{\pi}{2}\right)+\Gcal^{\otimes 3}_{k}\left(-\frac{\pi}{2}\right)\right).
\eea
For $\theta^{l}_{k}$ distributed according to the probability density of Eq.~\eqref{eq:haar_distribution_angle_i} we have (by identifying a ratio of Wallis integrals)
\bea
\expect{\sin^{2}(\theta^{l}_{k})} = \frac{k-1}{k}\,.
\eea
As a result, to generate a circuit uniformly over $\Mn^{+}\cap\Cliff$, it suffices to consider the matchgate circuits corresponding to the decomposition of Eq.~\eqref{eq:decomposition_in_rotations} and to independently sample each angle $\theta^{l}_{k}$ from the set of Clifford angles according to the distribution
\bea
p_{0} = p_{\pi} = \frac{1}{2k},\quad p_{\frac{\pi}{2}}=p_{-\frac{\pi}{2}} = \frac{k-1}{2k}\,.
\label{eq:distrib_four_cliff}
\eea

In the context of the classical shadows protocol, one can use the invariance under composition by reflections to further refine this distribution. Remarking that the Givens rotation $\Gcal_{k}(\pi)$ is equal to the composition of the reflections with respect to $\g_{k-1}$ and $\g_{k}$, we have $\Gcal_{k}(0)\sim\Gcal_{k}(\pi)$ and ${\Gcal_{k}(\pi/2)\sim\Gcal_{k}(-\pi/2)}$. Hence, for the shadows protocol one can simply use the decomposition of Eq.~\eqref{eq:decomposition_in_rotations} and sample each angle $\theta^{l}_{k}$ independently from $\{0,\pi/2\}$ according to the distribution
\bea
p_{0} = \frac{1}{k},\quad p_{\frac{\pi}{2}} = \frac{k-1}{k}\,.
\label{eq:distrib_two_cliff}
\eea

From the results of Sec.~\ref{sec:invariances_and_equivalences}, each rotation gate with an angle $\theta = \pm\pi/2$ corresponds, up to a reflection, to the application of a transposition. Therefore, the previous sampling scheme can be seen as a way to sequentially build a random permutation from independent random transpositions. Note that the structure of the circuit obtained by this sampling is not optimal in depth, due to inverted triangular shape that prevents an efficient parallelization. We provide a simple method to turn each circuit with the corresponding structure into an equivalent (up to reflections) more compact ``brick-wall'' structure in Appendix~\ref{app:brickwall_and_triangular_circuits_structures}. Unfortunately, this conversion does not allow to preserves the locality of the sampling, in the sense that the resulting random brick-wall circuit cannot be expressed as a product of local random gates. 

We investigate numerically and compare the performances of the shadows protocol associated with previous sampling methods to estimate the 1-RDM of a simple system of 8 qubits. Specifically, we consider random matchgates circuits with the triangular structure of Fig.~\ref{fig:matchgate_circuit_template_n=2} and random angles sampled according either to the distribution of Eq.~\eqref{eq:haar_distribution_angle_i} or to the distributions of Eqs.~\eqref{eq:distrib_two_cliff} and~\eqref{eq:distrib_four_cliff}. Figure~\ref{fig:numerical-results} shows the corresponding results. We use the absolute error for each of the observable estimators as measures of performance, and we plot the related relevant statistics averaged over the set of considered observables. As expected, the different sampling schemes yield shadow estimators with equivalent performances.  

The sampling methods considered so far yield circuits with a large number of gates. In view of obtaining an ensemble of matchgate circuits for the classical shadows protocol that are optimal in number of gates, we focus on sub-ensembles of $\Mn^{+}\cap\Cliff$ in the following. For the samplings corresponding to Eqs.~\eqref{eq:distrib_two_cliff} and~\eqref{eq:distrib_four_cliff}, one could try to reduce the number of gates by scanning the layers of the circuit and pruning the redundant gates. We leave the exploration of this path for future works. 

Here, we propose another method leveraging the results of Sec.~\ref{sec:invariances_and_equivalences}. Since the signs of the generalized permutations corresponding to the ensemble of Clifford matchgate circuits are irrelevant, one can first a draw a uniformly random permutation $P$. Then, we build the corresponding circuit by decomposing the permutation into a sequence of transposition and by using the equivalence between transpositions and Givens rotations with angle $\pm\pi/2$. Due to this equivalence, to obtain a minimal number of gates in the circuit amount to decompose $P$ into a minimal number of transpositions. The optimal decomposition can be computed using a bubblesort \cite{knuthArtComputerProgramming1998}, which will produce a circuit with a triangular structure and minimal number of gates. Using the method of Appendix~\ref{app:brickwall_and_triangular_circuits_structures}, we can finally to turn this circuit into an equivalent circuit with a brick-wall structure and the same number of gates. As the minimal number of transpositions required to decompose $P$ is equal to the number of its inversion~\cite{knuthArtComputerProgramming1998}, so is the number of gates in the associated circuit. As the expected number of inversion in a random permutation of the set $\left[2n\right]$ is equal to $n(n-1)/2$, a circuit obtained with this method will present the same expected number of gates. 

Eventually, as the shadows protocol is only sensitive to the perfect matching associated with the previous permutation, one can replace $P$ by any other permutation $P'$ satisfying $\mathrm{PerfMatch}(P)=\mathrm{PerfMatch}(P')$. To obtain a circuit with an optimal number of gates, we can chose the permutation $P'$ resulting in a minimal number of inversions. Given the perfect matching $\mathrm{PerfMatch}(P) = \{\{p_{2i-1}, p_{2i}\}, i\in[n]\}$, it suffices to consider the permutation obtained by first sorting each pair $\{p_{2i-1}, p_{2i}\}$ and then sorting the pairs by their first element. This clearly produces the permutation with the lowest number of inversion and thus the circuit with the lowest number of gates among the corresponding equivalence class. The average number of gates is then $n(n-1)/4$, with a maximal depth of $2n$ layers of commuting Givens rotations. This procedure is summarized in Algorithm~\ref{algo:sampling}, and the corresponding optimality result is encapsulated in the following proposition.

\begin{prop}
Algorithm~\ref{algo:sampling} produces circuits with the lowest number of gates among the class of circuits that are equivalent under the symmetries considered in Sec.~\ref{sec:invariances_and_equivalences}. Furthermore, the shadows protocol corresponding to the sub-ensemble of $\Mn^{+}\cap\Cliff$ generated by this algorithm is equivalent to the protocols associated with the ensembles of Table~\ref{tab:matchgate_ensembles}.
\end{prop}

Although this method is optimal in the number of gates, it is unclear whether it could be used to confirm the conjecture of Ref.~\cite{ogormanFermionicTomographyLearning2022}, as it would require a more detailed investigation of the average depth of the circuits obtained, which we leave for future work. 

\section{Conclusion}

In this paper, we investigated the matchgate classical shadows protocol associated with matrix group $\SO{2n}$, which remained unanalyzed in the previous related literature. Our approach relied on 
a decomposition of random matchgate circuits in products of independent Pauli rotations. By decomposing the quantum channels corresponding to these random rotations into convex sums of Clifford channels, we were able to show that the considered unitary ensemble admits the same first three moments as its intersection with the Clifford group. Thereby, we generalized a result that was previously known for the group of generalized matchgate circuits associated with $\Orth{2n}$. Extending existing results related to Clifford matchgate circuits, we further proved the equivalence between the shadows protocols corresponding to the various ensembles of matchgates studied in the literature. Building on our results, we also proposed new sampling schemes for different sub-ensembles of Clifford FGU, including a sampling scheme based on perfect-matching that is optimal in terms of number of gates.

We believe that our unifying results will prove very useful in
future applications, as they allow to transfer the results obtained for specific FGU ensembles to many others. In particular, our results show that one can use our gate-count optimal sampling scheme with the same performances guarantees as for the shadows protocol relying on the full ensemble of generalized FGUs, which is a clear improvement over the existing protocols. Along the way, we proved the existence of Clifford 3-cubatures for a large class of circuits of independent random Pauli rotations. We expect this result to find relevant applications in the contexts of randomized benchmarking and of variational quantum algorithms.

Many open questions and interesting research avenues remain. In this work, we derived a sampling method for the matchgate shadows protocol that is optimal in the number of gates. It would be interesting to analyze the scaling of the proposed scheme in terms of circuit depth. In particular, whether one could rely on the symmetry invariances of Sec.~\ref{sec:invariances_and_equivalences} to sample circuits with a proven minimal depth is unclear at this point. To further optimize the sampling scheme for matchgate shadows, a potentially fruitful research path would be to investigate the use of approximate matchgate design. This approach has recently proven successful in Ref.~\cite{schusterRandomUnitariesExtremely2024}, where the authors show that such an approximate ensemble can be used for Clifford shadows to obtain circuits with a logarithmic depth in the number of qubits. To obtain this approximate ensemble, one could for instance rely on classical random walks obtained from Markov chains applying random transpositions at each step. It is known that the mixing time of such Markov chain scales as $n^p\log(n)$~\cite{levinMarkovChainsMixing2017}, with $p=2$ for random transpositions of nearest neighbors and $p=1$ for transpositions between any pairs of the $n$ indices. Thus, depending on the connectivity of the considered device, it might possible to derive a sampling scheme yielding circuits with a sub-linear depth. However, it is unclear how to reliably build an estimator based on such an approximate ensemble, for the corresponding unitary ensemble would yields $t$-fold channels that would only approximate the ones of the matchgate ensemble. In the same spirit, and in view of the dependence of the shadows protocol on perfect-matchings, it would be worth exploring possible links with random phylo-genetic trees which can be seen as random perfect-matchings~\cite{holmesRandomWalksTrees2002}.

\section{Acknowledgments}

The authors acknowledge that this work was supported by Innovate UK - UKRI through grant 1468/10074167. We thank Thomas D. Barrett for his helpful suggestions during the writing of this paper.

\appendix

\section{Fermion-to-qubit mappings}
\label{app:jordan-wigner}
In this appendix, we review some facts about fermion-to-qubit mappings and recall the well-known Jordan-Wigner mapping. There exists a large variety of fermion-to-qubit mappings in the literature (see Refs.~\cite{bravyiFermionicQuantumComputation2002, jordanUeberPaulischeAequivalenzverbot1928,verstraeteMappingLocalHamiltonians2005, steudtnerFermiontoqubitMappingsVarying2018, jiangOptimalFermiontoqubitMapping2020, millerBonsaiAlgorithmGrow2023, chiewDiscoveringOptimalFermionqubit2023}, to cite but a few). Typically, such a mapping takes the form of a unitary transformation between the state-spaces of the considered fermionic and qubit systems. This transformation yields a homomorphism on the algebra of observables that preserves the algebraic properties of the operators. As the Majorana operators have the same algebraic properties as the Pauli operators, it is natural to require that the considered mapping sends Majorana operators to Pauli strings.

The Jordan-Wigner mapping identifies the fermionic Fock states $\ket{z_1\dots z_n}$ of the fermionic modes $a_1,\dots, a_n$ with the states $\otimes_{i=1}^{n}\ket{z_i}$ of canonical basis of the $\Hcal_{n}$, yielding the following correspondence between the Pauli and the mode operators:
\bea
X_k &= \left(\prod_{l<k}e^{i\pi \ad_{k}a_{k}}\right)\left(a^{\dagger}_k+a_k\right)\,,\\
Y_k &= \left(\prod_{l<k}e^{i\pi \ad_{k}a_{k}}\right)i\left(a^{\dagger}_k-a_k\right)\,,\\
Z_k &= 1-2a^{\dagger}_k a_k = e^{i\pi a^{\dagger}_k a_k}\,.
\eea
The mapping with the Majorana operators follows
\bea
\g_{2k-1} = \prod_{l<k}Z_l X_k\,,\quad
\g_{2k} = \prod_{l<k}Z_l Y_k\,,
\label{eq:Majorana_Pauli_map}
\eea
and states in the canonical basis can be written
\bea
\dyad{z} = \frac{1}{2^n}\prod_{k=1}^{n}
\left(I-i(-1)^{z_k}\g_{2k-1}\g_{2k}\right)\,.
\eea

\section{Proof of Lemma~\ref{lemma:3-fold_z_rotation}}
\label{app:proof_lemma_z_rotations}
In this appendix we prove Lemma~\ref{lemma:3-fold_z_rotation} of the main text. Recall that we denote 
\bea
{R}_{\theta} := e^{-i\frac{\theta}{2}Z} = e^{-i\frac{\theta}{2}}{\Pi}_{0}+e^{i\frac{\theta}{2}}{\Pi}_{1}\,,
\eea
with ${\Pi}_{0} := \dyad{0}$ and ${\Pi}_{1}:= \dyad{1}$, and that
\bea
{R}_{0} = I,\;{R}_{\frac{\pi}{2}} = S,\;{R}_{\pi} = Z,\; {R}_{\frac{3\pi}{2}} = S^{\dagger}\,.
\eea
The corresponding channels are respectively written $\Rcal_{\theta},\Ical, \Scal, \Zcal$ and $ \Scal^{\dagger}$. We restate the lemma for convenience.
\rotationlemma*
\textit{Proof:} For $1\leq k \leq 3$, define the projector onto the subspace of constant Hamming weight $k$
\bea
\Lambda_{k} := \sum_{\substack{k_1,k_2,k_3\in\{0,1\}\\ k_1+k_2+k_3=k}}{\Pi}_{k_1}\otimes{\Pi}_{k_2}\otimes{\Pi}_{k_3}\,.
\eea
We have
\bea
{R}_{\theta}^{\otimes 3} &= e^{-i\frac{3\theta}{2}}\Lambda_0+e^{-i\frac{\theta}{2}}\Lambda_1+e^{i\frac{\theta}{2}}\Lambda_2+e^{i\frac{3\theta}{2}}\Lambda_3 \\
&= e^{-i\frac{3\theta}{2}}\sum_{k=0}^{3} e^{ik\theta}
\Lambda_{k}\eea
such that
\bea
\Rcal_{\theta}^{\otimes 3}\left(\rho\right) = \sum_{k,l=0}^{3}e^{i\theta(k-l)}\Lambda_{k}\rho\Lambda_{l}\,.
\eea
For $\theta$ symmetrically distributed around $0$, we have $\expectt{\theta}{e^{i n\theta}} = \expectt{\theta}{e^{-i n\theta}}=\expectt{\theta}{\cos(n\theta)}$ for all ${n\in\mathbb{N}}$, and it comes
\bea
\expect{\Rcal_{\theta}^{\otimes 3}}\left(\rho\right) = \sum_{k,l=0}^{3} \expect{\cos(\theta(k-l))}\Lambda_{k}\rho\Lambda_{l}\,.
\eea
Using the parity of $\cos$ and setting $n=l-k$, we can reorder the sum to obtain
\bea
\expect{\Rcal_{\theta}^{\otimes 3}}\left(\rho\right) = \sum_{k=0}^{3}\sum_{n=0}^{3-k} \expect{\cos(\theta n)}\Ucal_{kn}\left(\rho\right)\,,
\label{eq:3-fold_symmetric}
\eea
where we defined
\bea
\Ucal_{kn}\left(\rho\right) := \Lambda_{k}\rhohat\Lambda_{k+n}+\Lambda_{k+n}\rho\Lambda_{k} - \delta_{n0}\Lambda_{k}\rhohat\Lambda_{k}\,.
\eea
Let $\delta(\theta)$ denote the Dirac distribution. Applying Eq.~\eqref{eq:3-fold_symmetric} for $\theta$ following the even distributions ${\delta(\theta-\pi)}$, ${\delta(\theta)}$ and ${\frac{1}{2}\left(\delta(\theta-\frac{\pi}{2})+\delta(\theta+\frac{\pi}{2})\right)}$ gives
\bea
\Zcal^{\otimes 3} &= \sum_{k=0}^{3}\sum_{n=0}^{3-k}(-1)^n\Ucal_{kn}\,,\\
\Ical^{\otimes 3} &= \sum_{k=0}^{3}\sum_{n=0}^{3-k}\Ucal_{kn}\,,\\
\frac{1}{2}\left(\Scal^{\otimes 3}+\Scal^{\dagger\otimes 3}\right) &= \sum_{k=0}^{3}\sum_{n=0}^{3-k} \left(\delta_{n0}-\delta_{n2}\right)\Ucal_{kn}\,.
\eea
Recombining the previous equations, we get
\bea
\frac{1}{4}\left(\Ical^{\otimes 3}+\Zcal^{\otimes 3}+\Scal^{\otimes 3}+\Scal^{\dagger\otimes 3}\right) &= \sum_{k=0}^{3}\sum_{n=0}^{3-k}\delta_{n0}\Ucal_{kn}\\
\frac{1}{4}\left(\Ical^{\otimes 3}+\Zcal^{\otimes 3}-\Scal^{\otimes 3}-\Scal^{\dagger\otimes 3}\right) &= \sum_{k=0}^{3}\sum_{n=0}^{3-k}\delta_{n2}\Ucal_{kn}
\label{eq:lc_3-fold_clifford_rotations}
\eea
Assuming further that $\theta$ is symmetrically distributed around $\frac{\pi}{2}$, and thus that $\theta$ is symmetric around every Clifford angle, we have that $\expect{\cos(n\theta)}=0$ for $n=1$ and $n=3$, but not necessarily for $n=2$. In that case, injecting Eq.~\eqref{eq:lc_3-fold_clifford_rotations} in Eq.~\eqref{eq:3-fold_symmetric} yields
\bea
\expect{\Rcal_{\theta}^{\otimes 3}} =& \frac{1}{4}\left(\Ical^{\otimes 3}+\Zcal^{\otimes 3}+\Scal^{\otimes 3}+\Scal^{\dagger\otimes 3}\right)\\ &+ \frac{\expect{\cos(2\theta)}}{4}\left(\Ical^{\otimes 3}+\Zcal^{\otimes 3}-\Scal^{\otimes 3}-\Scal^{\dagger\otimes 3}\right),
\eea
which gives the result outlined in Eq.~\eqref{eq:rotation-3-fold-channel}.

\section{Proof of Lemma~\ref{lemma:no-go-4-fold}}
\label{app:no-Clifford-4-cubature}

In this appendix we show that Lemma~\ref{lemma:3-fold_z_rotation} cannot be generalized to the $4$-fold channel by proving Lemma~\ref{lemma:no-go-4-fold}. As before, we write
\bea
R_{\theta} := \exp\left(-i\frac{\theta}{2}Z\right)\,
\eea
and $\Rcal_{\theta}$ the corresponding quantum channel.
Let us denote $\mathbb{U} = \left\{\Rcal_{\theta}, \theta\sim\nu\right\}$
the unitary ensemble associated with the random rotation angle $\theta$ distributed according to $\nu$. The distribution $\nu$ is not supported by the set of Clifford angles if it cannot be written as
\bea
\nu = \sum_{k=0}^{3}p_k \delta_{k\pi/2}\,
\eea
with $p_k \geq 0$ and $\sum_{k=0}^{3}p_k = 1$.
We restate Lemma~\ref{lemma:no-go-4-fold} for convenience.

\nofourfoldlemma*

\textit{Proof:} Suppose that $\mathbb{U}$ admits a Clifford $4$-cubature $\left\{ (\Ccal_j, p_j)\,,\,j\in J\right\}$ such that
\bea
\expect{\Rcal_{\theta}^{\otimes 4}} = \sum_{j\in J}p_j \Ccal_{j}^{\otimes 4}\,.
\eea
For every $j\in J$, there is a unique Pauli operator ${P_j\in\{I,X,Y,Z\}}$ satisfying $\Ccal_{j}(X) = \lambda_j P_j$ with $\abs{\lambda_j}=1$. We denote
\bea
\Gamma :=  \sum_{P\in\left\{I,X,Y,Z\right\}^{\otimes 4}}\frac{1}{2^4}\left\vert\Tr\left\{P\expect{\Rcal_{\theta}\left(X\right)^{\otimes 4}}\right\}\right\vert\,,
\eea
and from the above we have
\begin{align}
\begin{split}
\\
\Gamma =& \sum_{P\in\left\{I,X,Y,Z\right\}^{\otimes 4}}\left\vert\sum_{j\in J} \frac{p_j}{2^4}\Tr\left\{P\Ccal_{j}(X)^{\otimes 4}\right\}\right\vert\\
&\leq \sum_{P\in\left\{I,X,Y,Z\right\}^{\otimes 4}}\sum_{j\in J} \frac{p_j}{2^4}\left\vert\Tr\left\{P P_{j}^{\otimes 4}\right\}\right\vert\\
&\leq \sum_{j\in J} p_j = 1\,.
\end{split}
\label{eq:sum_pauli_contributions_4_fold}
\end{align}
On the other hand, for all $\theta\in[-\pi,\pi)$ we have
\bea
\Rcal_{\theta}(X) = \cos(\theta)X+\sin(\theta)Y
\eea
such that
\bea
\expect{\Rcal_{\theta}(X)^{\otimes 4}} =\expect{\left(\cos(\theta)X+\sin(\theta)Y\right)^{\otimes 4}}\,.
\eea
Developing this equation, we obtain
\begin{widetext}
\bea
\expect{\Rcal_{\theta}(X)^{\otimes 4}} =&\; \expect{\cos(\theta)^4}XXXX+\expect{\sin(\theta)^4}YYYY \\
&+ \expect{\cos(\theta)\sin(\theta)^3}\left(XYYY+YXYY+YYXY+YYYX\right)\\
&+ \expect{\cos(\theta)^{3}\sin(\theta)}\left(YXXX+XYXX+XXYX+XXXY\right)\\
&+ \expect{\cos(\theta)^{2}\sin(\theta)^{2}}\left(XXYY+YYXX+XYXY+YXYX+XYYX+YXXY\right)\,,
\eea
\end{widetext}
where we dropped the tensor products for clarity.
Assuming that $\theta$ is symmetrically distributed about the Clifford angles, we have
\begin{align}
\expect{\cos(\theta)\sin(\theta)^3}=\expect{\cos(\theta)^{3}\sin(\theta)}=0\,,
\end{align}
which gives
\bea
\Gamma &= \expect{\cos(\theta)^4}+\expect{\sin(\theta)^4}+6\expect{\cos(\theta)^2\sin(\theta)^2}\\
&= 1 + 4\expect{\cos(\theta)^2\sin(\theta)^2} \,.
\eea
Assuming further that $\nu$ is not supported by the set of Clifford angles, we have $\expect{\cos(\theta)^2\sin(\theta)^2}>0$ such that $\Gamma > 1$.
This contradicts Eq.~\eqref{eq:sum_pauli_contributions_4_fold}, hence $\mathbb{U}$ do not admit any Clifford $4$-cubature.

\section{Distribution of the angles of Givens rotations}
\label{app:euler_angles_distrib_adaptation}

In this appendix, we provide details and justify our adaptation of the Proposition 1.6. of Ref.~\cite{meckesRandomMatrixTheory2019}. Up to a change of notations, the original version of the proposition proves the result for following distribution for the independent rotations angles
\bea
\theta^{l}_{2} \sim \frac{d\theta^{l}_{2}}{2\pi}\quad\text{on}\quad[0,2\pi)
\eea
and $\forall k \in (2, 2n]$,
\bea
\theta^{l}_{k}\sim \frac{\Gamma\left(\frac{k-1}{2}\right)}{2\Gamma\left(\frac{1}{2}\right)\Gamma\left(\frac{k-2}{2}\right)}\sin(\theta)^{k-2}\quad\text{on}\quad[0,\pi)\,.
\eea

The proof of Ref.~\cite{meckesRandomMatrixTheory2019} relies on a colmun-by-column construction of the random rotation matrix $Q$. Before we review this construction, we recall some facts related to random vectors on spheres. Let $\mathbb{S}^{n-1}\subset \mathbb{R}^{n}$ be the $(n-1)$-sphere, a vector $\vb{v}\in\mathbb{S}^{n-1}$ is characterized by its associated Euler angles $\theta_1,\dots, \theta_{n-1}$ as follow
\bea
\vb{v} = \begin{pmatrix}
\sin(\theta_{n-1})\dots\sin(\theta_{2})\sin(\theta_1)\\
\sin(\theta_{n-1})\dots\sin(\theta_{2})\cos(\theta_1)\\
\vdots\\
\sin(\theta_{n-1})\cos(\theta_{n-2})\\
\cos(\theta_{n-1})
\end{pmatrix}\,,
\label{eq:euler-angles}
\eea
with $ 0\leq \theta_{1} < 2\pi$ and $0\leq \theta_{k}\leq\pi$ for ${2 \leq k\geq n-1}$. For $\vb{v}$ to be uniformly distributed over $\mathbb{S}^{n-1}$, it suffices that its Euler angles are distributed according to the measure
\bea
\d\mu_{\mathbb{S}^{n-1}} := \frac{\Gamma\left(\frac{n}{2}\right)}{(2\pi)^{n/2}}\left(\prod_{k=2}^{n}\sin^{k-1}(\theta_{k})\boldsymbol{1}_{[0,\pi)}(\theta_{k})\d\theta_{k}\right)\d\theta_{1}\,,
\label{eq:uniform-measure-sphere}
\eea
where $\boldsymbol{1}_{A}$ is the indicator function of the set $A$. 

Remark that in the previous definition of the Euler angles, the domains of the $\theta_{k}$ for ${2\leq k \leq n-1}$ can be freely chosen to be either $[0,\pi]$ or $[-\pi, 0]$ without loss of generality. We use this freedom to extend the domain of the random Euler angles associated with uniformly distributed random vectors. 

Let $\vb{v}(\theta_1,\dots,\theta_{n-1})$ be defined by Eq.~\eqref{eq:euler-angles} with $\theta_{1},\dots\theta_{n-1}$ distributed according to the measure of Eq.~\eqref{eq:uniform-measure-sphere}. Since $\vb{v}$ is uniformly distributed, so is $Q'\vb{v}$ for any matrix $Q'\in\Orth{n}$. Take $Q'$ to be the reflection with respect to the $i$-th axis and let $X$ be a Bernouilli random variable with ${\mathbb{P}(X=1)=\mathbb{P}(X=0)=1/2}$, we have that
\bea
\vb{u} := (1-X)\vb{v} + X Q'\vb{v}\,
\eea
is also uniformly distributed on $\mathbb{S}^{n-1}$, with
\bea
Q'\vb{v} = \vb{v}(\theta_1,\dots, -\theta_{i}\,.\dots\theta_{n-1}).
\eea
As a result, replacing
\bea
\sin^{i-1}(\theta_{i})\boldsymbol{1}_{[0,\pi)}(\theta_{i})
\eea
by 
\bea
\frac{1}{2}\abs{\sin(\theta_{i})}^{i-1} = \frac{1}{2}\Bigl(&\sin^{i-1}(\theta_{i})\boldsymbol{1}_{[0,\pi)]}(\theta_{i})
\\
&+\sin^{i-1}(-\theta_{i})\boldsymbol{1}_{(-\pi,0]}(\theta_{i})\Bigr)
\eea
in the probability distribution of Eq.~\eqref{eq:uniform-measure-sphere} above still yields a uniformly distributed vector. As this holds for any index $i$, we get that any vector defined by Eq.~\eqref{eq:euler-angles} such that $\theta_1,\dots \theta_{n-1}$ follow the distribution
\bea
\d\tilde{\mu}_{\mathbb{S}^{n-1}} := \frac{\Gamma\left(\frac{n}{2}\right)}{2^{n-1}(2\pi)^{n/2}}\left(\prod_{k=2}^{n}\abs{\sin(\theta_{k})}^{k-1}\d\theta_{k}\right)\d\theta_{1}\,
\label{eq:uniform-measure-sphere-extended}
\eea
is uniformly distributed on $\mathbb{S}^{n-1}$.

The constrution of a random element $Q\in\SO{2n}$ presented in Ref.~\cite{meckesRandomMatrixTheory2019} then proceeds as follow. We write $\vb{q}_i$ the $i$-th column of the matrix $Q$. The first step of the construction is to draw the last column of $Q$, i.e. $\vb{q}_{2n} = Q\vb{e}_{2n}$, uniformly from the sphere $\mathbb{S}^{2n-1}$. To do so, one can take $\vb{q}_{2n} = Q_1\vb{e}_{2n}$ with $Q_1 := g^{1}_{2}g^{1}_{3}\dots g^{1}_{2n}$. This vector has the form given in Eq.~\eqref{eq:euler-angles} and the vector Euler angles are the ones of the rotations $g^{1}_{k}$. From the above, choosing these angles randomly according the distribution of Eq.~\eqref{eq:uniform-measure-sphere-extended} yields a vector uniformly distributed on $\mathbb{S}^{2n-1}$. Then, the second-to-last column $\vb{q}_{2n-1} = Q\vb{e}_{2n-1}$ is uniformly sampled in the $(2n-2)$-sphere of the orthogonal complement of $Q\vb{e}_{2n}$, namely $\mathbb{S}^{2n-2}\cap \{Q\vb{e}_{2n}\}^{\perp}$. As before, it suffices to take the vector $Q_1 Q_2 \vb{e}_{2n-1}$ with $Q_2 = g^{2}_{2}g^{2}_{3}\dots g^{2}_{2n-1}$ and to sample the angles of $Q_2$ from the distribution of Eq.~\eqref{eq:uniform-measure-sphere-extended}. Multiplying on the left by $Q_1$ allows to sample the resulting vector from the orthogonal complement of $\vb{q}_{2n}$, as we have
\bea
\langle \vb{q}_{2n}, \vb{q}_{2n-1} \rangle &= \langle \vb{e}_{2n} Q_{1}^{T}Q_{1}Q_{2}\vb{e}_{2n-1}\rangle \\
&= \langle \vb{e}_{2n} Q_{2}\vb{e}_{2n-1}\rangle \\
&= 0\,.
\eea

Proceeding like that up to the first column yields a Haar-distributed random matrix $Q$ of the form given in the main text, that is
\bea
    Q = (g^{1}_{2}g^{1}_{3}\dots g^{1}_{2n})\dots(g^{2n-2}_{2}g^{2n-2}_{3})(g^{2n-1}_{2})\,.
\eea
Importantly, for each $1\leq k\leq 2n-1$ the angles of the Givens rotations $g^{k}_{l}(\theta^{k}_{l})$ correspond to the random Euler angles of the column vector $Q\vb{e}_{2n+1-k}$. In particular, in the proof of Ref.~\cite{meckesRandomMatrixTheory2019}, the distribution of the angles is inherited from the probability distribution of Eq.~\eqref{eq:uniform-measure-sphere} used to generated the various random vectors.
Consequently, the proof remains valid if we replace the distribution of Eq.~\eqref{eq:uniform-measure-sphere} by the one of Eq.~\eqref{eq:uniform-measure-sphere-extended} and our adapted version of the proposition follows.

\section{Brick-wall and triangular circuits structures}
\label{app:brickwall_and_triangular_circuits_structures}

In this appendix, we give a simple method to transform circuits with a triangular structure described in the sampling scheme of Sec.~\ref{sec:sampling-schemes} into a circuit with a ``brick-wall'' structure that is equivalent for the classical shadows protocol. 

\begin{figure*}
    \includegraphics[width=\textwidth]{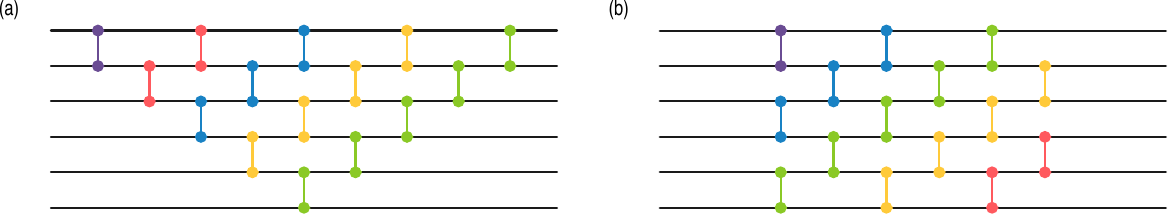}
    \caption{Structures of the sequences of transposition considered in this work: (a) triangular structure and (b) brick-wall structure. Each line corresponds to a single Majorana operator and connections between adjacent lines correspond to the application of either the identity or to the transposition of the two lines.}
    \label{fig:brickwall-triangular-structures}
\end{figure*}
\begin{figure*}
    \includegraphics[width=\textwidth]{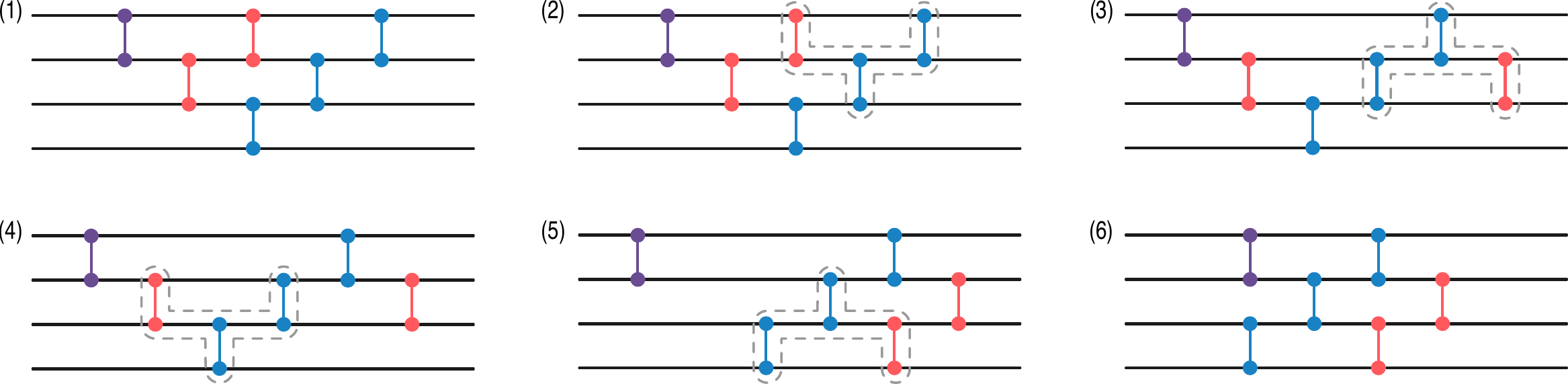}
    \caption{Example of elementary transformations allowing to turn a triangular structure into a brick-wall one. Subfigures (1) and (6) respectively show the intial and target structure. Subfigures (2) and (3) (resp. (4) and (5)) show the first (resp. second) steps of the the transformation. The outlined sets of connections represents equivalent groups of transpositions that are mapped to each other through Eq.~\eqref{eq:equi-transpo}. Generalizing and repeating this sequence of operations allows to exchange the order of the diagonals of different length in the structures of Fig.~\ref{fig:brickwall-triangular-structures}, which in turns allows to map triangular structures to brick-wall ones.}
    \label{fig:triangular-to-brickwall-transform}
\end{figure*}

We consider matchgate circuits that can be decomposed according to Eq.~\eqref{eq:decomposition_in_rotations} with angles belonging to $\{0, \pi/2\}$. The corresponding quantum circuit is given for $n=1$ qubit in Fig.~\ref{fig:matchgate_circuit_template_n=2} for the JW mapping. Recall that under this mapping, each Pauli rotation corresponds to a Givens rotation acting on a pair of adjacent indices. From the invariance results of Sec.~\ref{sec:invariances_and_equivalences}, each Givens rotation of angle $\pi/2$ acting on the pair of indices $(i,i+1)$ corresponds to a transposition of these indices. Thus, one can represent the equivalence class of circuits yielding the same permutation by a sequence of transposition. For the considered circuits, the obtained sequence is represented by a triangular ``circuit'' of transpositions as represented on the left panel of Fig.~\ref{fig:brickwall-triangular-structures}.

The triangular shape is inherited from the decomposition of Eq.~\eqref{eq:decomposition_in_rotations}. There exists other such decomposition in the literature. In particular, \citet{clementsOptimalDesignUniversal2016} provide an analog decomposition leading to a circuit of Givens rotations arranged in a rectangular structure, which we refer to as a ``brick-wall'' structure. Under the previous equivalence, this decomposition leads to circuits of transposition with the shape given on the right panel of Fig.~\ref{fig:brickwall-triangular-structures}.

In order to turn circuit in $\Mn\cap\Cliff$ obtained from the sampling scheme of the Sec.~\ref{sec:sampling-schemes} into an equivalent circuit with a brick-wall structure, it suffices to transform its corresponding triangular circuit of transposition into a brick-wall one and chose a quantum circuit in the associated equivalence class. 

To transform a triangular circuit of transpositions into a brick-wall one, we propose a simple strategy that consists in permuting the successive diagonals of transpositions as represented on Fig.~\ref{fig:brickwall-triangular-structures}. Denote $\tau_i$ the transposition of the indices $(i,i+1)$. It is well known that such transpositions satisfy the following braid relation
\bea
\tau_i\tau_{i+1}\tau_i = \tau_{i+1}\tau_i\tau_{i+1}\,.
\eea
Let $b\in\{0,1\}$, and write $\tau_i^{b}$ the permutation equal to $\tau_i$ if $b=1$ and that is the identity otherwise. A simple inspection shows that for all $b_1, b_2, b_3 \in \{0,1\}$ there exists $b'_1, b'_2, b'_3 \in \{0,1\}$ such that
\begin{equation}
\begin{split}
&\tau^{b_1}_i\tau^{b_2}_{i+1}\tau^{b_3}_i = \tau^{b'_1}_{i+1}\tau^{b'_2}_{i}\tau^{b'_3}_{i+1}\,,\\
&b_1+b_2+b_3 \geq b'_1+b'_2+b'_3\,.
\end{split}
\label{eq:equi-transpo}
\end{equation}
Figure~\ref{fig:triangular-to-brickwall-transform} shows how the previous relation can be used to permute two diagonals. This is straightforwardly extended to diagonals of any size.

\end{document}